\documentclass[conference]{IEEEtran}

\usepackage{amsmath,amssymb}
\usepackage{todonotes}
\usepackage{url}% http://ctan.org/pkg/url
\usepackage{amsmath}
\usepackage{algpseudocode}
\usepackage{algorithm}
\usepackage{hhline}
\usepackage{graphicx}
\graphicspath{ {fig/} }
\usepackage{svg}
\usepackage{subfig}
\usepackage{float}
\usepackage{adjustbox}
\usepackage{tabularx}
\begin{document}
\title{\textbf{PowerHammer:} Exfiltrating Data from Air-Gapped Computers through Power Lines}

\author{\IEEEauthorblockN{Mordechai Guri, Boris Zadov, Dima Bykhovsky, Yuval Elovici\IEEEauthorrefmark{3}}
	\IEEEauthorblockA{Ben-Gurion University of the Negev, Israel\\Cyber-Security Research Center\\\IEEEauthorrefmark{3}Department of Software and Information Systems Engineering\\ Email: gurim@post.bgu.ac.il, borisza@gmail.com, bykhov@post.bgu.ac.il, elovici@bgu.ac.il}}

\maketitle

\begin{abstract}
In this paper we provide an implementation, evaluation, and analysis of \textit{PowerHammer}, a malware (bridgeware \cite{Guri:2018:BAM:3200906.3177230}) that uses power lines to exfiltrate data from air-gapped computers. In this case, a malicious code running on a compromised computer can control the power consumption of the system by intentionally regulating the CPU utilization. Data is modulated, encoded, and transmitted on top of the current flow fluctuations, and then it is conducted and propagated through the power lines. This phenomena is known as a 'conducted emission'. We present two versions of the attack. \textit{Line level power-hammering}: In this attack, the attacker taps the in-home power lines\footnote{Commonly referred to as \textit{electrical wiring} or \textit{household electricity mains}. this paper they will be referred to simply as \textit{power lines}} that are directly attached to the electrical outlet. \textit{Phase level power-hammering}: In this attack, the attacker taps the power lines at the phase level, in the main electrical service panel. In both versions of the attack, the attacker measures the emission conducted and then decodes the exfiltrated data. We describe the adversarial attack model and present modulations and encoding schemes along with a transmission protocol. We evaluate the covert channel in different scenarios and discuss signal-to-noise (SNR), signal processing, and forms of interference. We also present a set of defensive countermeasures. Our results show that binary data can be covertly exfiltrated from air-gapped computers through the power lines at bit rates of 1000 bit/sec for the line level power-hammering attack and 10 bit/sec for the phase level power-hammering attack.

\end{abstract}


% \fi
\IEEEpeerreviewmaketitle

\section{Introduction}
Information is the most critical asset of modern organizations, and accordingly it is coveted by adversaries. Protecting IT networks from sophisticated cyber-attacks is quite a complicated task, involving host level and network level security layers. This includes updating protection software in the host computers, configuring firewalls and routers, managing access controls, using centralized credential systems, and so on. Nevertheless, despite a high degree of protection, as long as the local area network has a connection with the outside world (e.g., the Internet), irrespective of the level or type of protection, an innovative and persistent attacker will eventually find a way to breach the network, eavesdrop, and transmit sensitive data outward (e.g., see the Vault 7 \cite{macaskill2017wikileaks}, Sony \cite{zetter2014sony}, and Yahoo \cite{thielman2016yahoo} incidents).

\subsection{Air-Gapped Networks}
When sensitive data is involved, an organization may resort to so-called 'air-gap' isolation. An air-gapped network is a secured computer network in which measures are taken to maintain both physical and logical separation from less secured networks. The air-gap separation is maintained by enforcing strict regulations such as prohibiting connectivity to unauthorized equipment and hardening the workstations in the network. Today, air-gapped networks are used in military and defense systems, critical infrastructure, the finance sector, and other industries \cite{guri2017bridging,byres2013air}. Two examples of air-gapped networks are the United States’ Defense Intelligence Agency’s  NSANET and Joint Worldwide Intelligence Communications System (JWICS) classified networks \cite{Classifi75:online}.
However, even air-gapped networks are not immune to breaches. In the past decade, it has been shown that attackers can successfully penetrate air-gapped networks by using complex attack vectors, such as supply chain attacks, malicious insiders, and social engineering \cite{maybury2005analysis,TrumpPut87:online,abraham2010overview}. In 2017, WikiLeaks published a reference to a hacking tool dubbed 'Brutal Kangaroo' used to infiltrate air-gapped computers via USB drives \cite{Wikileak92:online}. This tool was used by the attackers to infect the Internet workstations of an organization's employees, and wait for an employee to insert the infected USB drive into an air-gapped computer. Using such tools, attackers can breach the network, bypassing security systems such as AVs, firewalls, intrusion detection and prevention systems (IDS/IPS), and the like. 

\subsection{Air-Gap Covert Channels}
After deploying a malware in the air-gapped network, the attacker may, at some point, wish to leak information - a behavior commonly used in advanced persistent threats (APTs). However, despite the fact that breaching air-gapped systems has been shown feasible, the exfiltration of data from an air-gapped system remains a challenge. Over the years, various out-of-band communication methods to leak data through air-gaps have been proposed. For example, electromagnetic covert channels have been studied for at least twenty years. In this type of communication, a malware modulates binary information over the electromagnetic waves radiating from computer components (LCD screens, communication cables, computer buses, and hardware peripherals \cite{guri2014airhopper,kuhn1998soft,kuhn2002compromising,vuagnoux2009compromising,guri2015gsmem}). Other types of air-gap covert channels based on magnetic \cite{guri2018odini},\cite{guri2018magneto}, acoustic \cite{hanspach2014covert}, \cite{guri2017acoustic} optical, \cite{Guri2017} and thermal \cite{guri2015bitwhisper} emissions have also been investigated.

\subsection{Exfiltration through Power Lines}
In this paper, we present a new type of electric (current flow) covert channel.  
The method, dubbed \textit{PowerHammer}, enables attackers to exfiltrate information from air-gapped networks through AC power lines. We show that a malware running on a computer can regulate the power consumption of the system by controlling the workload of the CPU. Binary data can be modulated on the changes of the current flow, propagated through the power lines, and intercepted by an attacker. We present two versions of the attack. In \textit{line level power-hammering} the adversary taps the power cables feeding the transmitting computer. In \textit{phase level power-hammering} the adversary taps the power lines at the phase level, in the main electrical service panel. Using a non-invasive tap, the attacker measures the emission conducted on the power cables. Based on the signal received, the transmitted data is demodulated and decoded back to a binary form.  
\\

The rest of this paper is structured as follows: Technical background is provided in Section \ref{sec:tech}. Related work is presented in Section \ref{sec:related}. The adversarial attack model is discussed in Section \ref{sec:attack}. Communication, modulation and protocol details are described in Section \ref{sec:communication}. Section \ref{sec:eval} provide the analysis and evaluation results. Countermeasures are discussed in Section \ref{sec:counter}. We conclude the paper in Section \ref{sec:conclusion}.

\section{Technical Background}\label{sec:tech}

\begin{figure}
	\centering
	\includegraphics[width=\linewidth]{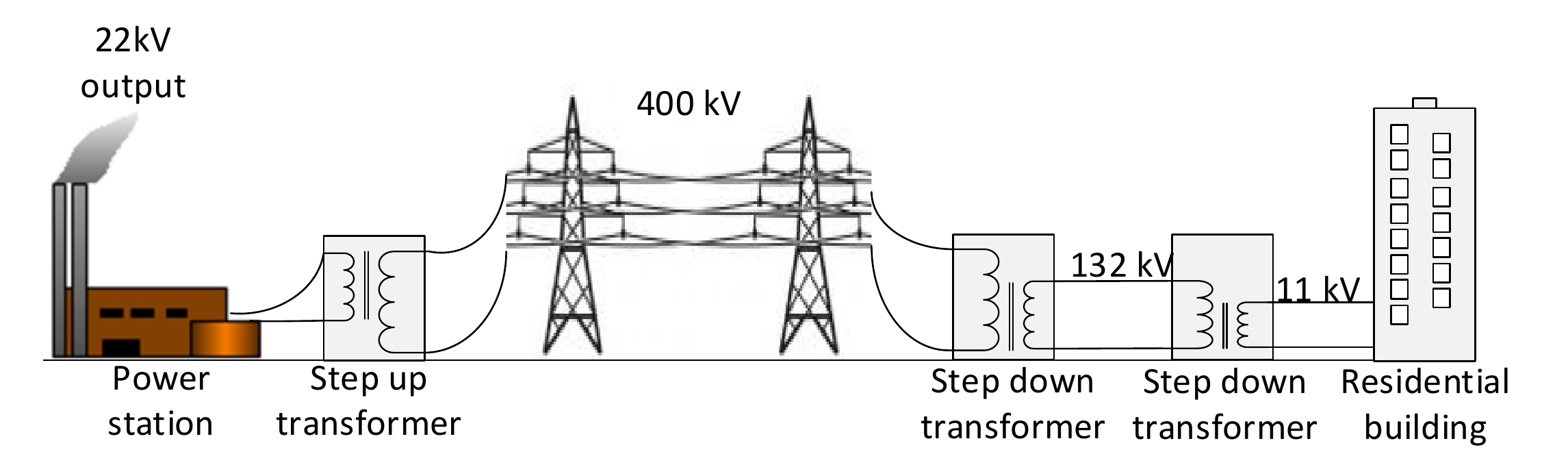}
	\caption{A (simplified) schematic diagram of the electricity network}
	\label{fig:elecnetwork}
\end{figure}

\subsection{Electricity Network}
The electrical grid refers to the electrical network which supplies electricity to the end consumers. The electricity is supplied to the consumers (e.g., industrial buildings, homes, farms, etc.) through power lines. Fig.~\ref{fig:elecnetwork} illustrates a simplified electricity network, all the way from the power station to an end consumer. The power station  generates electricity from steam, by burning fossils such as coal or natural gas. The voltage increases up to 400kV using step-up transformers and is transmitted through high voltage power lines to the city's power station. From city stations, the electricity is supplied to large buildings (e.g., at 11kV) and small residential buildings (e.g., at 220v).

\subsubsection{Phases}
\begin{figure}
	\centering
	\includegraphics[width=\linewidth]{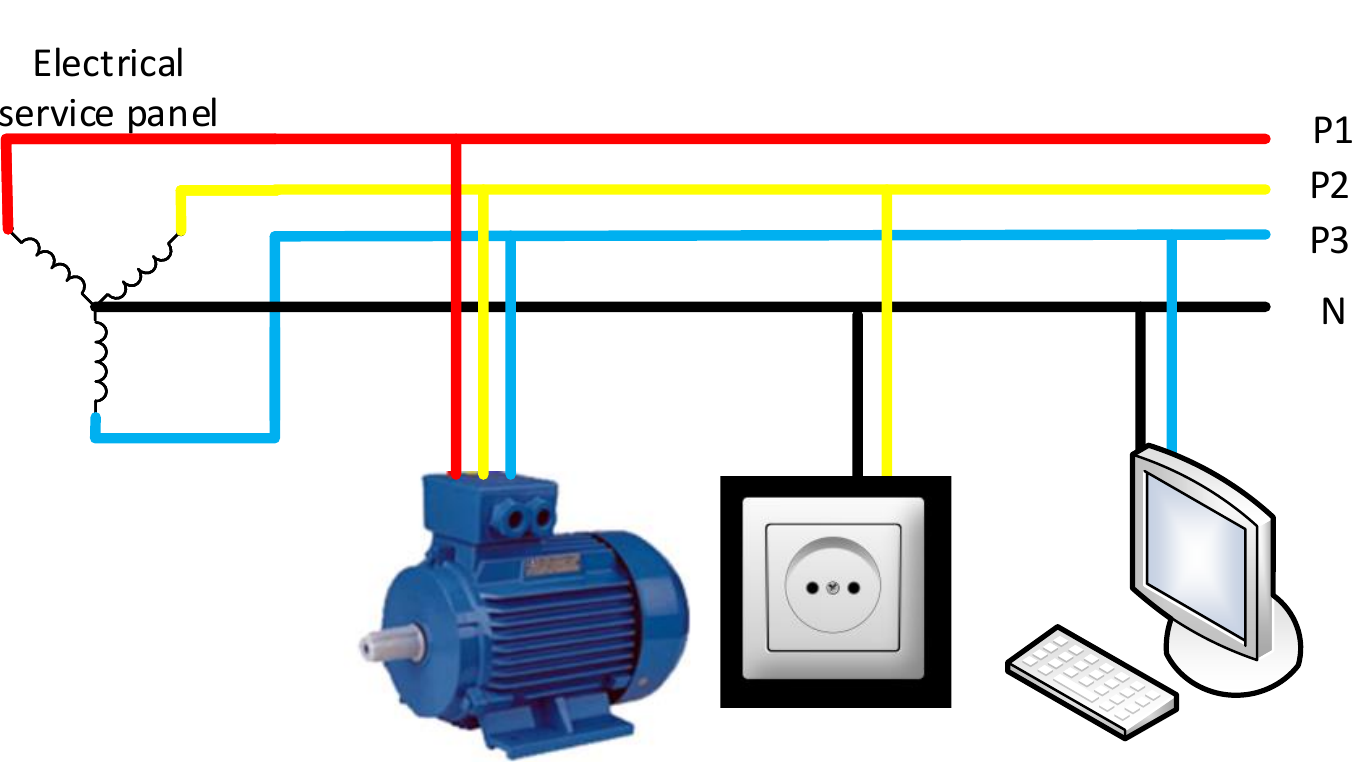}
	\caption{Phase interconnection in the home main electric panel}
	\label{fig:Phase1}
\end{figure}

The typical in-door electrical supply network is comprised of a  distribution board (controlled from the main electrical service panel) that divides electrical power into subsidiary phases. Each phase feeds several to dozens of circuits. An illustration of such a network is presented in Fig.~\ref{fig:Phase1}. The distribution to three different phases prevents internal wall wires from overheating, and allow to run fairly large loads.
 
\subsubsection{Power line distribution in building} 
\begin{figure}
	\centering
	\includegraphics[width=\linewidth]{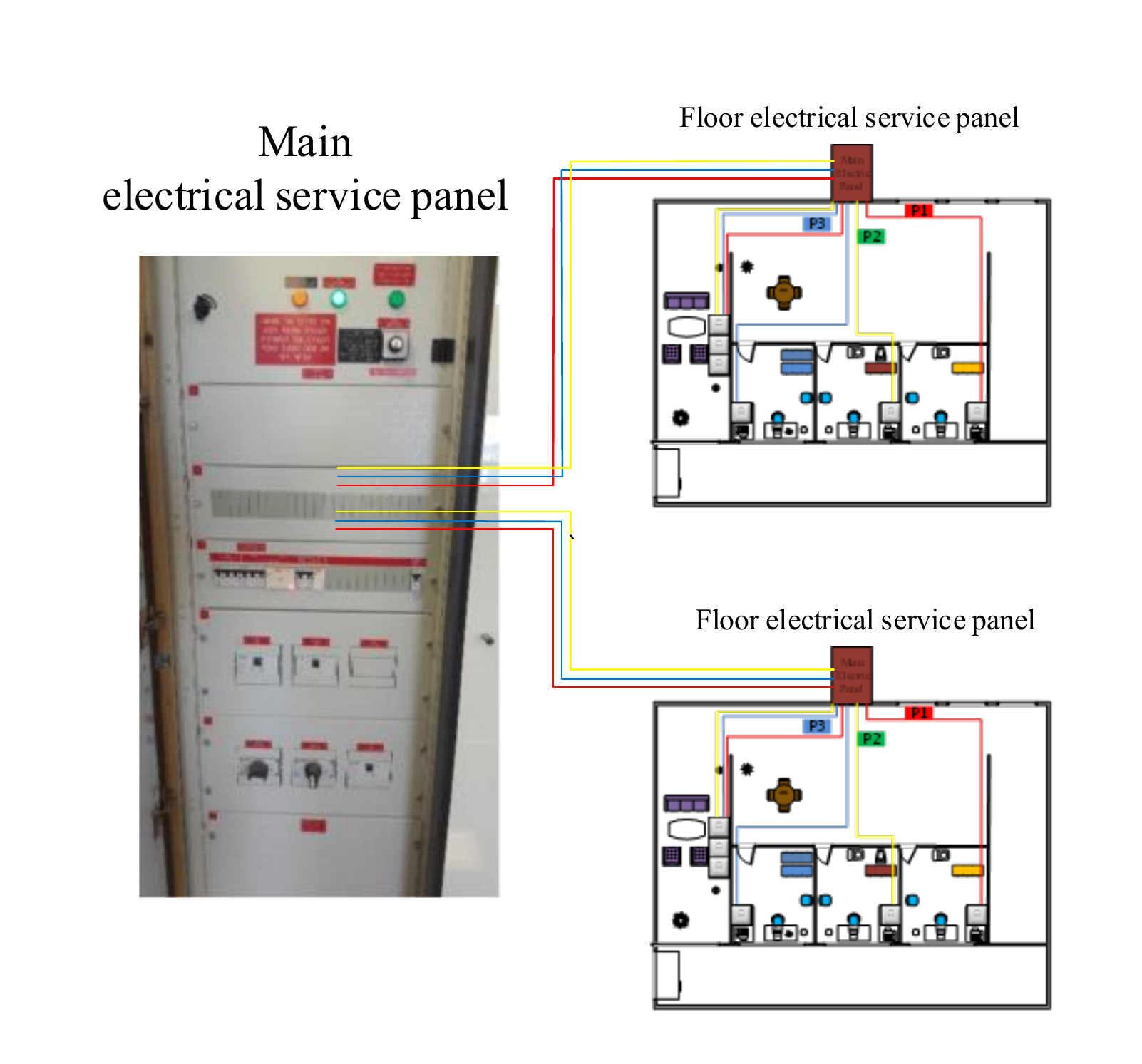}
	\caption{In-door power distribution}
	\label{fig:building_distribution}
\end{figure}

Electricity distribution in buildings is managed in a main, centralized electrical service panel (Fig. \ref{fig:building_distribution}). Every floor in the building has a dedicated electrical service panel, which is connected to the main service panel of the building. The power cords are connected to the floor panel with circuit breakers.

\subsection{Switch-Mode Power Supplies}
 Computers consume power using their power supplies. Today, modern switch-mode power supplies (SMPS) are used in all types of modern electronic equipment, including computers, TVs and mobile phones. The main advantages of SMPSs over the older linear power supplies include  their higher efficiency, smaller size, and lighter weight. Technically, an SMPS transfers power from a 220V AC source to several DC loads while converting voltage and current characteristics. Briefly, in the SMPS, the AC power passes through fuses and a line filter, and then is rectified by a full-wave bridge rectifier. Next, the rectified voltage is applied to the power factor correction (PFC) pre-regulator followed by DC to DC converters. An in-depth discussion on the design of SMPSs is beyond the scope of this paper, and we refer the interested reader to handbooks on this topic      \cite{handbookbillings2011switchmode}.

 \subsubsection{Power Line Conducted Emission} 
 The SMPS generates electromagnetic radiation due to the fast switching current and voltage they generate during the normal operation \cite{handbookbillings2011switchmode,counterliu2002high,counterye2004novel}. The radiated noise is also leaked to the power lines due to electromagnetic conductive processes, known as conduction emission \cite{fiori2003comparison,Conducte52:online}.  As investigated in many works in this field, the source of the electromagnetic and conductive noise is rooted in the internal transistors and their switching states \cite{handbookbillings2011switchmode,SMPSSMPSRMre91:online,counterliu2002high}. The switching generates AC voltage and current flows through the internal circuits which produces electromagnetic emissions. Although these emissions can be limited with proper design and shielding, there is still interference from high frequency transformers in the power supply 
 \cite{handbookbillings2011switchmode,counterliu2002high}. 
 
 Note that there are various regulations on the permitted levels of the radiated and conducted emissions from electronic devices.  Commercially available power supplies are designed to meet these regulations. The FCC Part 15 regulations requires that the conducted emission be controlled at the frequency bands of 450kHz-30MHz (or 150kHz-30MHz)  \cite{Microsof90:online}\cite{Introduc8:online}. Our covert channel uses frequencies lower than 24kHz, and hence works even with computers that are fully complaint with these regulations. Studies on the topic of electromagnetic and conductive noise and their implications can be found in \cite{fiori2003comparison,Conducte52:online}.
Our method exploits the conducted emission to modulate binary information on the power lines. We show that a malicious code can influence the momentary power consumption of the computer, generating data-modulated conduction on the power lines in the low frequency band. The generated noise travels along the input power lines and can be measured by an attacker probing the power cables.

\section{Related Work}
\label{sec:related}
Air-gap covert channels are special covert channels that enable communication from air-gapped computers - mainly for the purpose of data exfiltration. In 2018, researcher Mordechai Guri coinded the term \textit{bridgeware} \cite{Guri:2018:BAM:3200906.3177230} to refer to the class of malware that exploits air-gap covert channels in order to bridge the air-gap between isolated computers/ networks and attackers. The covert channels can be classified into five main categories: electromagnetic, magnetic, acoustic, thermal, and optical. This paper introduce the electric (current flow) based covert channel.

\subsection{Electromangetic}
In the past twenty years, various studies have explored the use of electromagnetic emanation for covert communication. Back in 1998, Kuhn et al showed that it is possible to regulate the electromagnetic emissions from  computer display cables \cite{kuhn1998soft}. They showed that it is possible to control the frequencies of the generated electromagnetic waves precisely and modulate information on top of them. Based on this work, Thiele \cite{Tempestf48:online} presented an open-source program (Tempest for Eliza) which uses the computer monitor to transmit AM radio signals modulated with audio. 
He demonstrated the method by transmitting the tune Beethoven's tune, Letter to Elise, and showed how it could be heard from a simple radio receiver located nearby. More recently, Guri et al demonstrated AirHopper \cite{guri2014airhopper,guri2017bridging}, a malware that exfiltrates data from air-gapped computers to a nearby smartphone via radio signals emitted from the screen cable. The covert transmissions are received by the FM radio receiver which is integrated into many modern smartphones. In 2015, Guri et al demonstrated GSMem \cite{guri2015gsmem}, malware that leaks data from air-gapped computers to nearby mobile phones. They successfully generated frequencies in the cellular bands (in the GSM, UMTS, and LTE specifications) from the buses which connect the RAM and the CPU. The transmission received by a rootkit hidden in the baseband processor of a nearby mobile-phone. In 2016, Guri et al presented USBee, a malware that uses the USB data buses to generate electromagnetic signals from a desktop computer \cite{guri2016usbee}. The transmissions could be picked up by a simple RF receiver in a radius of several meters. Similarly, researchers also proposed using GPIO ports of printers to generate covert radio signals for the purpose of data exfiltration \cite{funtenna86:online}.
 
\subsection{Magnetic}
In 2018, Guri et al presented ODINI \cite{guri2018odini}, a malware that can exfiltrate data from air-gapped computers via low frequency magnetic signals generated by the computer's CPU cores. They showed that the low-frequency magnetic fields bypass Faraday cages and metal shields. Guri et al also demonstrated MAGNETO \cite{guri2018magneto}, a malware which is capable of leaking data from air-gapped computers to nearby smartphones via magnetic signals. They used the magnetometer sensor integrated in smartphones to measure the change in magnetic fields. Matyunin suggested using the head of magnetic hard disk drives (the actuator arm) to generate magnetic emission, which can be received by a nearby smartphone magnetic sensor \cite{matyunin2016covert}.

\subsection{Optical}
Several studies have proposed the use of optical emanation from computers for covert communication. Loughry introduced the use of PC keyboard LEDs (caps-lock, num-lock and scroll-lock) to encode binary data \cite{loughry2002information}. The main
drawback of this method is that it is not completely
covert. Since keyboard LEDs don't typically
flash, the user may detect the communication. In 2017, Guri et al presented LED-it-GO, a covert channel that uses the hard drive (HDD) indicator LED in order to exfiltrate data from air-gapped computers \cite{Guri2017}. Guri et al also presented a method for data exfiltration from air-gapped networks via router and switch LEDs \cite{guri2017xled}. In the case of HDDs and routers, the devices blinks frequently, and hence, the transmissions performed via these channels will not raise the user's suspicious. Research showed that data can also be leaked optically through fast blinking images or low contrast invisible bitmaps projected on the LCD screen \cite{guri2016optical}.  In 2017, Guri et al presented aIR-Jumper, a malware that uses security cameras and their IR LEDs to covertly communicate with air-gapped networks from a distance of hundreds of meters \cite{guri2017air}. Similarly, Lopes presented a hardware based approach for leaking data using infrared LEDs maliciously installed on a storage device. Note that their approach requires the attacker to deploy the compromised hardware in the organization. 

\subsection{Thermal}
In 2015, Guri et al introduced  BitWhisper \cite{guri2015bitwhisper}, a thermal covert channel allowing an attacker to establish bidirectional communication between two adjacent air-gapped computers via temperature changes. The heat is generated by the CPU/GPU of a standard computer and received by temperature sensors that are integrated into the motherboard of the nearby computer.

\subsection{Acoustic}
In acoustic covert channels, data is transmitted via inaudible, ultrasonic sound waves. Audio based communication between computers was reviewed by Madhavapeddy et al in 2005 \cite{madhavapeddy2005audio}. In 2013, Hanspach \cite{hanspach2014covert} used inaudible sound to establish a covert channel between air-gapped laptops equipped with speakers and microphones. Their botnet established communication between two computers located ~19 meters apart and can achieve a bit rate of 20 bit/sec. Deshotels \cite{deshotels2014inaudible} demonstrated the acoustic covert channel with smartphones, and showed that data can be transferred up to 30 meters away. The aforementioned ultrasonic covert channels are relevant to environments in which the computers are  equipped with both speakers and microphones, a less common setup in many secure environments. To overcome this limitation, Guri et al presented MOSQUITO \cite{Guri18Mosquito} a malware that covertly turns headphones, earphones, or simple earbuds connected to a PC into a pair of  microphones, even when a standard microphone is muted, taped, turned off or not present. Using this technique they established so-called speaker-to-speaker ultrasonic communication between two or more computers in the same room.

In 2016, Guri et al introduced Fansmitter, a malware which facilitates the exfiltration of data from an air-gapped computer via noise intentionally emitted from PC fans \cite{guri2016fansmitter}. In this method, the transmitting computer does not need to be equipped with audio hardware or an internal or external speaker. Guri et al also presented DiskFiltration a method that uses the acoustic signals emitted from the hard disk drive (HDD) moving arm to exfiltrate data from air-gapped computers \cite{guri2017acoustic}.

\begin{table}[]
	\centering
	\caption{Summary of existing air-gap covert channels}
	\renewcommand{\arraystretch}{1.2}	
	\label{table-related}
	\begin{tabular}{ll}
		\hline
		Type               & Method                                                                                                                                                                                          \\ \hhline{==}
		Electromagnetic    & \begin{tabular}[c]{@{}l@{}}AirHopper \cite{guri2014airhopper,guri2017bridging} (FM radio) \\ GSMem \cite{guri2015gsmem} (cellular frequencies)  \\ USBee \cite{guri2016usbee} (USB bus emission) \\ Funthenna \cite{funtenna86:online} (GPIO emission) \end{tabular}                                                                   \\ \hline
		Magnetic           & \begin{tabular}[c]{@{}l@{}}MAGNETO \cite{guri2018magneto} (CPU-generated \\ magnetic fields)\\ ODINI \cite{guri2018odini} (Faraday shields bypass) \\ Hard-disk-drive \cite{matyunin2016covert} \end{tabular}            
		\\ \hline
		Electric (current)           & \begin{tabular}[c]{@{}l@{}}PowerHammer (this paper) \end{tabular}                                  \\                                      \hline
		Acoustic           & \begin{tabular}[c]{@{}l@{}}Fansmitter \cite{guri2016fansmitter} (computer fan noise) \\ DiskFiltration \cite{guri2017acoustic} (hard disk noise) \\ Ultrasonic \cite{hanspach2014covert,carrara2014acoustic}\\ MOSQUITO \cite{Guri18Mosquito} (speaker-to-speaker \\ultrasonic communication)\end{tabular} \\ \hline \\
		Thermal            & BitWhisper  \cite{guri2015bitwhisper} (heat emission)                                                                                                                                                                           \\\\ \hline 
		Optical            & \begin{tabular}[c]{@{}l@{}}LED-it-GO \cite{Guri2017} (hard drive LED) \\ VisiSploit \cite{guri2016optical} (invisible pixels) \\ Keyboard LEDs \cite{loughry2002information}\\ Router LEDs \cite{guri2017xled}\end{tabular}                            \\ \hline 
		Optical (infrared) & aIR-Jumper \cite{guri2017air} (security cameras)                                                        \\ \hline
	\end{tabular}
\end{table}

\subsection{Electric (Power Consumption)}
In this work we introduce the use of power lines to exfiltrate data from air-gapped computers. Note that utilizing power cables for carrying data is well known technique, usually referred to as power-line communication (PLC) \cite{ferreira2010power}. Today, the PLC technologies are used for a wide range of industrial and home applications, including high-speed Internet, machine-to-machine communication, smart homes  and so on. However, all current PLC methods focus on a non-malicious type of communication. In addition, they require special modems for the transmission and reception, which aim at filtering and modulating/demodulating the information carried on the power lines. We explore the use of power lines for a \textit{malicious} covert communication. We generate parasitic signals on the power lines, using a malware running on a standard computer, without the need for any additional hardware components for transmission.

Table \ref{table-related}. summarizes the existing air-gap covert channels.

\begin{figure}
	\centering
	\includegraphics[width=0.9\linewidth]{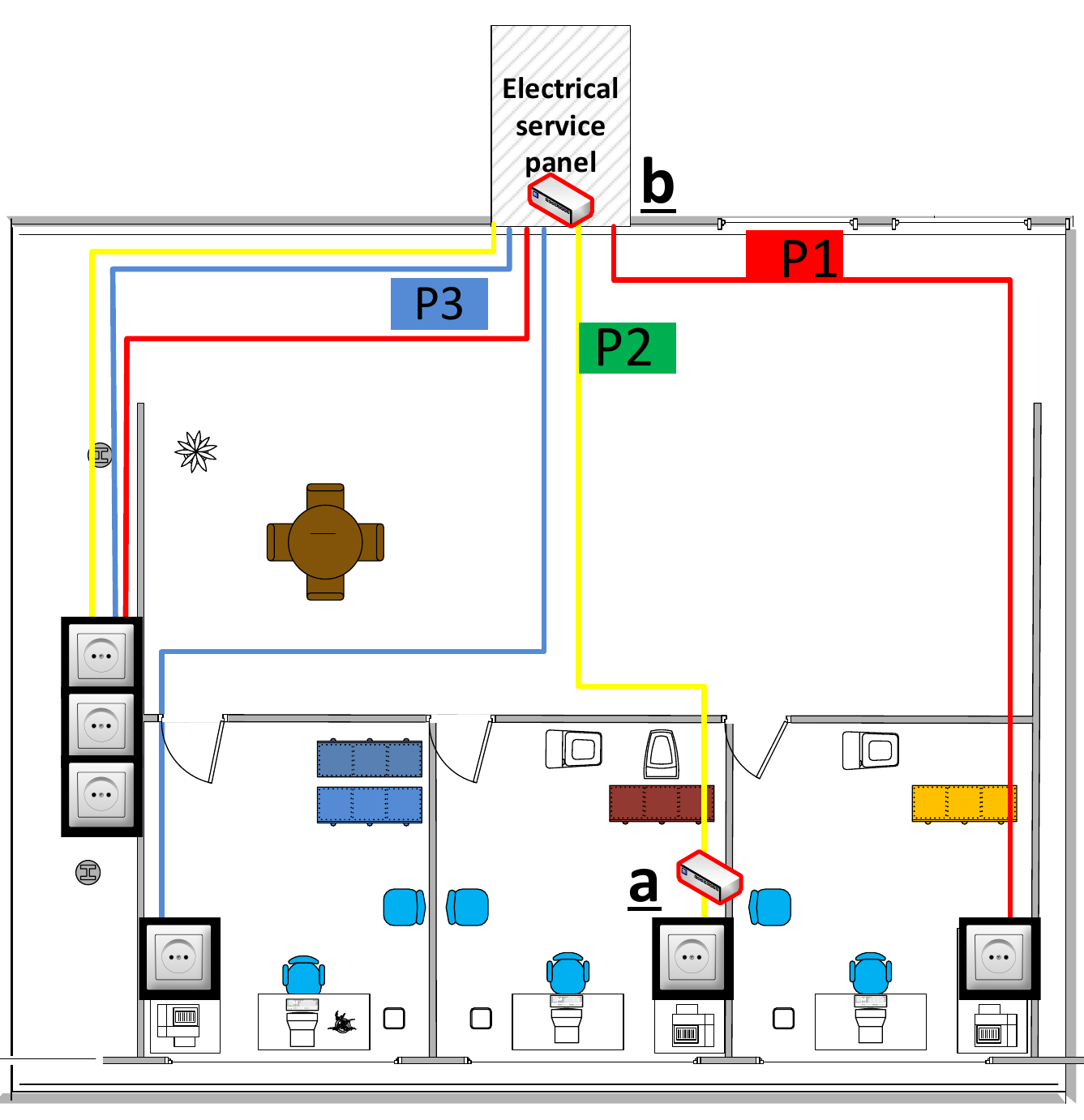}
	\caption{The receiver implantation: (a) on the power line feeding the computer, and (b) on the power line in the main electrical service panel} 
	\label{fig:room}
\end{figure}

\section{Adversarial Attack Model}
\label{sec:attack}
The attack consists of four main steps: (1) system infection, (2) receiver implantation, (3) data gathering, and (4) data exfiltration.
 
\subsection{System Infection} 
The adversarial attack model requires running a malicious code in the targeted air-gapped computer. 
In the incursion phase, the attacker infects the target system or network with malware. Infecting highly secure and even air-gapped networks has been proven feasible in recent years. Note that several APTs discovered in the last decade are capable of infecting air-gapped networks \cite{AndreyNi62:online}, e.g., Turla \cite{TheEpicT20:online}, RedOctober \cite{zaored}, and Fanny \cite{AFannyEq68:online}. As a part of the targeted attack, the adversary may infiltrate the air-gapped networks using social engineering, supply chain attacks, or malicious insiders.

\subsection{Receiver Implantation}
The receiver is a non-invasive probe connected to a small computer (for the signal processing). The probe is attached to the power line feeding the computer (Fig. \ref{fig:room}a) or the main electric panel (Fig. \ref{fig:room}b). It measures the current in the power line, process the modulated signals, decodes the data and sends it to the attacker (e.g., with Wi-Fi transceiver). An example of such a hardware implant was presented in Snowden's leaked documents. In that case, the component, known as COTTONMOUTH \cite{COTTONMO81:online}, was a USB connector with hidden RF transceiver that attackers used to connect with air-gapped networks by sending and receiving data to/from a long haul relay subsystem .  
 
\subsection{Data Gathering}
Having a foothold in the system, the malware starts retrieving interesting data for the attacker. The data might be files, encryption keys, credential tokens, or passwords. The data gathered is exfiltrated through a computer, usually a PC workstation or server which contains the sensitive data to leak. 

\subsection{Exfiltration}
In the last phase of the attack, the malware starts the data leak by encoding the data and transmitting it via signals injected to the power lines. The signals are generated by changing the workload on the CPU cores. The transmissions may take place at predefined times or in response to some trigger infiltrated by the attacker. The signal is received by the power line probe and delivered to the attacker (e.g., via Wi-Fi).
\\

Note that although the described attack model is complicated, it is not beyond the ability of motivated and capable attackers. Advanced persistent threats coupled with sophisticated attack vectors such as supply chain attacks and human engineering have been shown to be feasible in the last decade. As a reward for these efforts, the attacker can get his/her hands on very valuable and secured information, which is out of reach of other types of covert channels.

\section{Communication}
\label{sec:communication}
In this section we describe the signal generation algorithm and present the data modulation schemes and the transmission protocol.

\subsection{Signal Generation}
As described in the scientific background section, changes in the power consumption of a computer can be measured on the power lines.  In a standard computer, the current flows primarily from wires that supply electricity from the main power supply to the motherboard. The CPU is one of the greatest consumers of power in the motherboard. Since modern CPUs are energy efficient, the momentary workload of the CPU directly affects the dynamic changes in its power consumption \cite{von2016variations}. By regulating the workload of the CPU, it is possible to govern its power consumption, and hence control the current flow in the cable. In the most elementary case, overloading the CPU with calculations will consume more current. Intentionally starting and stopping the CPU workload allows us to generate a signal on the power lines at specified frequency and modulate binary data over it. 
We developed a fine-grained approach, in which we control the workload of each of the CPU cores independently from the other cores. Regulating the workload of each core separately enables greater control of the momentary power consumption. This approach has two main advantages: 

\begin{enumerate}
	\item Choosing which cores to operate on at a given time, allows us to use only the currently available cores, that is, cores which are not utilized by other processes. This way, the transmission activity won’t interrupt other active processes in the system. This is important for the usability of the computer and the stealth of the covert channel.
	
	\item By using different numbers of cores for the transmission, we can control the current consumption (e.g., fewer cores consume less power), and hence the amplitude of the carrier wave. This allows us to employ amplitude based modulations in which data is encoded on the amplitude level of the signal. 	
\end{enumerate}

To generate a carrier wave at frequency $f_c$ in one or more cores, we control the utilization of the CPU at a frequency correlated to $f_c$. To that end, $n$ worker threads are created where each thread is bound to a specific core. To generate the carrier wave, each worker thread overloads its core at a frequency $f_c$ repeatedly – alternating between applying a continuous workload on its core for a time period of $1/2f_c$ (full power consumption) and putting its core in an idle state for a time period of $1/2f_c$ (no power consumption).

\begin{algorithm} % enter the algorithm environment
	\caption{WorkerThread ($iCore$, $freq$, $nCycles0$, $nCycles1$)} % give the algorithm a caption
	\label{alg1} % and a label for \ref{} commands later in the document
	\begin{algorithmic}[1] % enter the algorithmic environment
		\State $ bindThreadToCore(iCore) $
		\State $half\_cycle\_ms \gets 0.5*1000/freq $
		
		\While {$ (!endTransmission()) $}
		\If {$ (data[i] = 0)  $}
		\State sleep($nCycles0$*half\_cycle\_ms*2)
		\Else
		\For{$j \gets 0\ to\ nCycles1$}
		\State $ T1 \gets getCurrentTime()$
		\While{$ (getCurrentTime()-T1 < half\_cycle\_ms)$} \texttt{;}
		\EndWhile
		\State {$ sleep(half\_cycle\_ms) $ }
		\EndFor
		\EndIf
		\EndWhile
		
	\end{algorithmic}
\end{algorithm}

\begin{figure}
	\centering
	\includegraphics[width=0.8\linewidth]{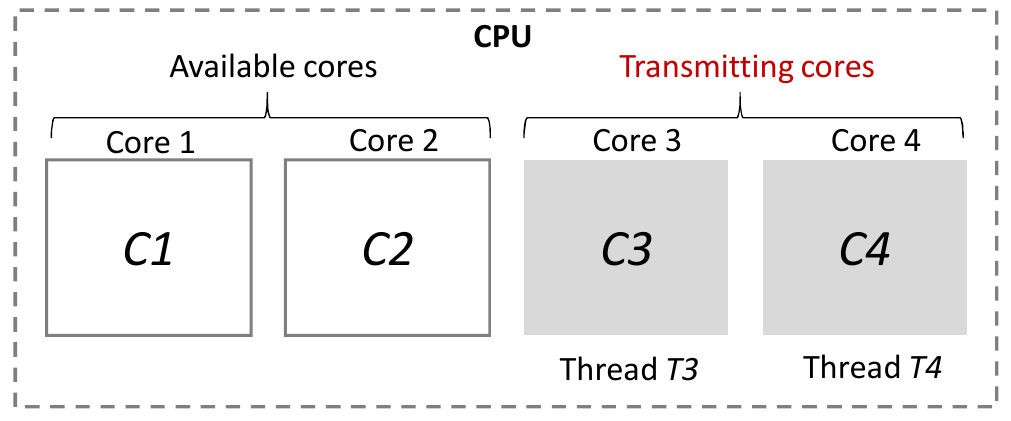}
	\caption{A CPU with two transmitting threads}
	\label{fig:cores}
\end{figure}

This operation is illustrated in Figure \ref{fig:cores}, which depicts a system with two worker threads. Threads $T3$ and $T4$ are bound to cores $C3$ and $C4$, respectively. Note that cores $C1$ and $C2$ don’t participate in this transmission. When the worker threads $T3$ and $T4$ start, they receive the required carrier frequency $f_c$ and the stream of bits to transmit. The basic operation of a worker thread is described in algorithm 1.

A worker thread receives the core to be bound to ($iCore$) and the carrier frequency ($freq$). It also receives the number of cycles for the modulation of logical '0' ($nCycles0$) and the number of cycles for the modulation of logical '1' ($nCycles1$). Note that the cycle time is derived from the frequency of the carrier wave (line 2). The thread’s main function iterates on the array of bits to transmit. In the case of logical '0' it sleeps for $nCycles0$ cycles (line 5). In the case of logical '1' it repeatedly starts and stops the workload of the core at the carrier frequency $freq$ for $nCycles1$ cycles (lines 8-10). We overload the core using the busy waiting technique. This causes full utilization of the core for the time period and returns. 

Based on the algorithm above, we implemented a transmitter for Linux Ubuntu (version 16.04, 64 bit). We used the \texttt{sched\underline{ }setaffinity} system call to bind each thread to a CPU core. The affinity is the thread level attribute that is configured independently for each worker thread. To synchronize the initiation and termination of the worker threads, we used the thread mutex objects with \texttt{pthread$\_$mutex$\_$lock()} and \texttt{pthread$\_$mutex$\_$unlock()} \cite{pthreadm53:online}. For thread sleeping we used the \texttt{sleep()} system call \cite{sleep3Li5:online}. Note that the precision of \texttt{sleep()} is in milliseconds, and it is sufficient given the frequencies of the carrier waves which are  at 24kHz or lower. 
%communication section

\subsection{Data Modulation \& Communication Signals}
\begin{figure}[h]
	\centering
	\includegraphics[width=0.85\linewidth,page=3]{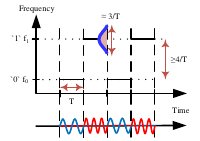}
	\caption{Illustration of FSK modulation}
	\label{fig:fsk_illustration2}
\end{figure}

Recall that the transmitting code can determine the frequency of the signal by setting the cycle time in the signal generation algorithm. In the current research we used frequency shift keying (FSK) for data modulation. In FSK the data is represented by a change in the frequency of a carrier wave. 
For the evaluation we used the binary frequency-shift keying (B-FSK) modulation \cite{Proakis2007} outlined in Fig. \ref{fig:fsk_illustration2}.
In this modulation, only two different carrier frequencies are employed. Each frequency is amplitude modulated, such that '1' and '0' are two possible symbol combinations. The generalized form of this modulation (M-FSK), is evaluated later, in Section \ref{sec:eval}.
In FSK, the length of each symbol is $T$, and the spectral width is approximately ${\pm 1.5/T}$. Theoretically, this modulation requires at least ${4/T}$ frequency spacing between different carrier frequencies.

\begin{figure}[t]
	\centering
	\includegraphics[width=\linewidth,page=5]{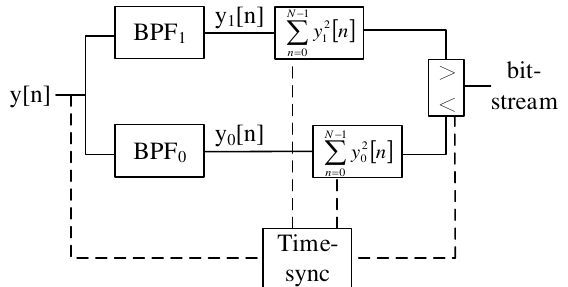}
	\caption{Block scheme of receiver}
	\label{fig:receiver}
\end{figure}

The main advantage of the proposed modulation that it does not require channel state information (CSI) when a \textit{non-coherent} receiver is applied \cite{Proakis2007}. Such a receiver (Fig. \ref{fig:receiver}) inludes the band-pass filters (BPFs) BPF\textsubscript{0} and BPF\textsubscript{1} that are located around frequencies $f_0$ and $f_1$ respectively. The energy of $y_0[n]$ and $y_1[n]$ signals is compared over a time period of $T$ and the received bit is determined by the highest energy value.

\begin{figure}[t]
	\centering
	\includegraphics[width=\linewidth]{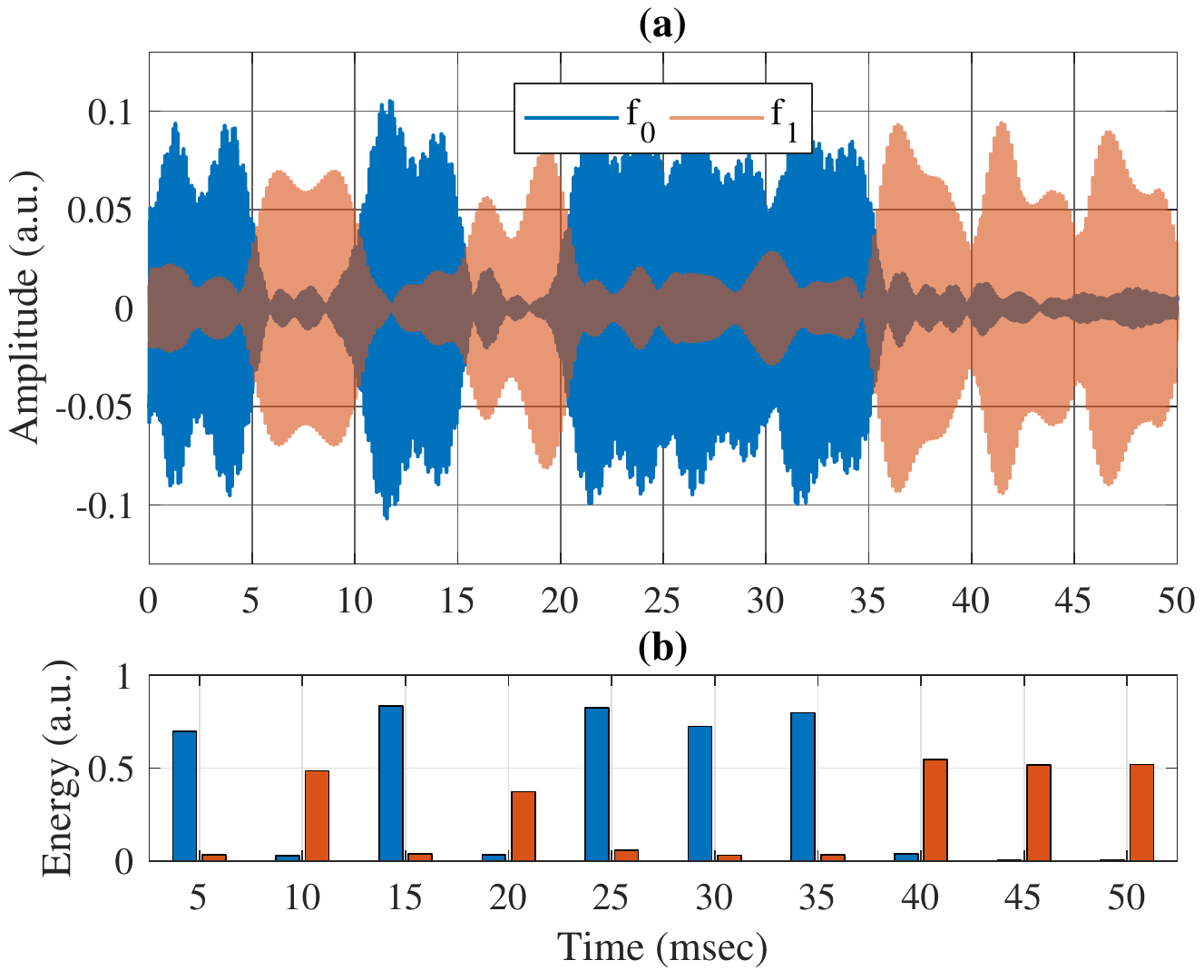}
	\caption{Time domain plot of a communication signal with a symbol length of 5 msec (bit rate of 200 bit/sec) and communication frequencies ${f_0=10}$kHz and ${f_1=18}$kHz}
	\label{fig:plottime1}
\end{figure}

In the following paragraphs, we discuss some of the results regarding the B-FSK modulation.
Fig. \ref{fig:receiver} presents the block scheme of a receiver. A sample signal is presented in Fig. \ref{fig:plottime1}.
In this transmission (a sequence of '0101000111') the symbol's time is $T=5$ msec which yields a bit rate of 200 bit/sec. For each symbol, energies of $y_0[n]$ and $y_1[n]$ are compared, as outlined in Fig. \ref{fig:plottime1}(b).

\begin{figure}
	\centering
	\includegraphics[width=0.8\linewidth]{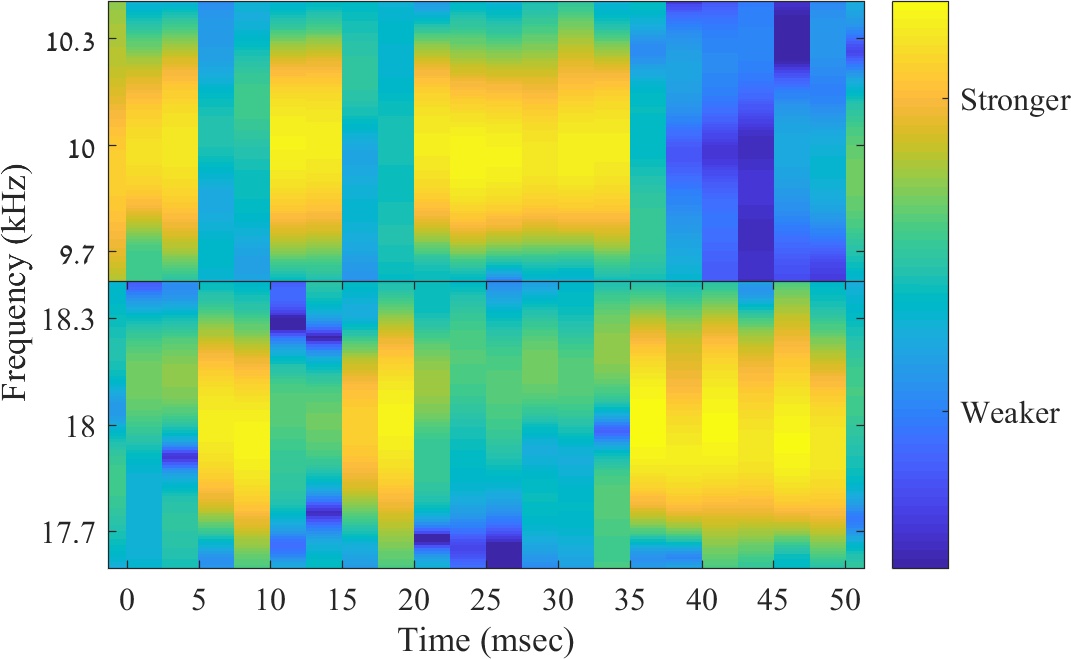}
	\caption{Spectrogram of B-FSK communication at 200 bit/sec; the transmitted signal is the '0101000111' sequence}
	\label{fig:plotspectrogram}
\end{figure}

Fig.~\ref{fig:plotspectrogram} shows the spectrogram (also termed as short-time Fourier transform) of the '0101000111' sequence. In this case, $T=5$ msec which yields a bit rate of 200 bit/sec. We used the two frequencies ${f_0=10}$kHz and ${f_1=18}$kHz for the B-FSK modulation. The time spacing of the spectrogram was chosen to match the symbol length of two bins per symbol. The spectral width of the signals matches the theoretical bandwidth of ${\cong 3/T = 600}$Hz. 

\subsection{Bit-Framing}
The data packets are transmitted in small frames. Each
frame consists of 44 bits and is comprised of a preamble,
payload, and CRC (cyclic redundancy check).

\begin{itemize}
	\item	\textbf{Preamble.} Like most air-gap covert channels, the unidirectional communication means that the receiver cannot establish a handshake with the transmitter, and hence cannot determine or set the channel parameters before the transmission. To solve this, a preamble header is transmitted at the beginning of every packet. It consists of a sequence of four alternating bits ('1010') which helps the receiver determine the properties of the channel, such as the carrier wave frequency and amplitude. In addition, the preamble allows the receiver to detect the beginning of a transmission and synchronize itself.  
	\item \textbf{Payload.} The payload is the raw data to be transmitted. In our case, we arbitrarily choose 32 bits as the payload size. 
	\item \textbf{CRC.} For error detection, we insert eight bits of CRC code at the end of the frame. The receiver calculates the CRC for the received payload, and if it differs from the received CRC, an error is detected. 
\end{itemize}

\section{Analysis \& Evaluation}
\label{sec:eval}
In this section we provide the analysis and evaluation of the PowerHammer attack. We present the experimental setup and discuss the measurements and results.

\subsection{Experimental Setup}

\subsubsection{Measurement setup}\label{sec:measurement_setup}
\begin{figure}
	\centering
	\includegraphics[width=\linewidth]{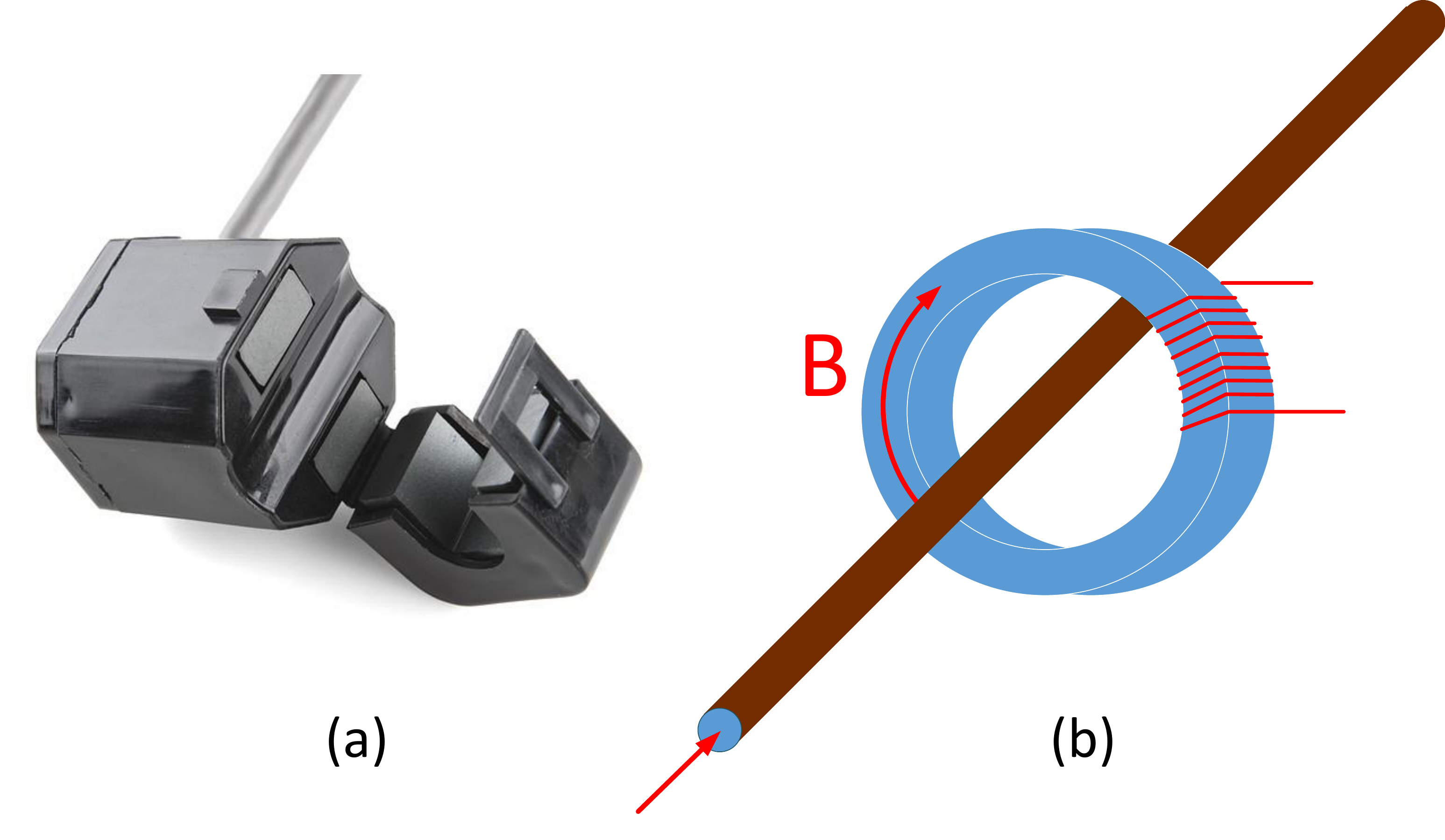}
	\caption{A split core current transformer}
	\label{fig:transformer-schematics}
\end{figure}

Our measurement setup consists of a current probe connected to a measurement system. For measuring the current, we used a type of split core current transformer. This is a non-invasive probe which is clamped around the power line and measures the amount of current passing through it (Fig. \ref{fig:transformer-schematics}). The non-invasive probe behaves like an inductor which responds to the magnetic field around a current-carrying cable (Fig. \ref{fig:transformer-schematics} b). The amount of current in the coil is correlated with the amount of current flowing in the conductor. For our experiments we used SparkFun's split core current transformer ECS1030-L72 \cite{Microsof94:online}. 

\begin{figure}[h]
	\centering
	\includegraphics[width=\linewidth]{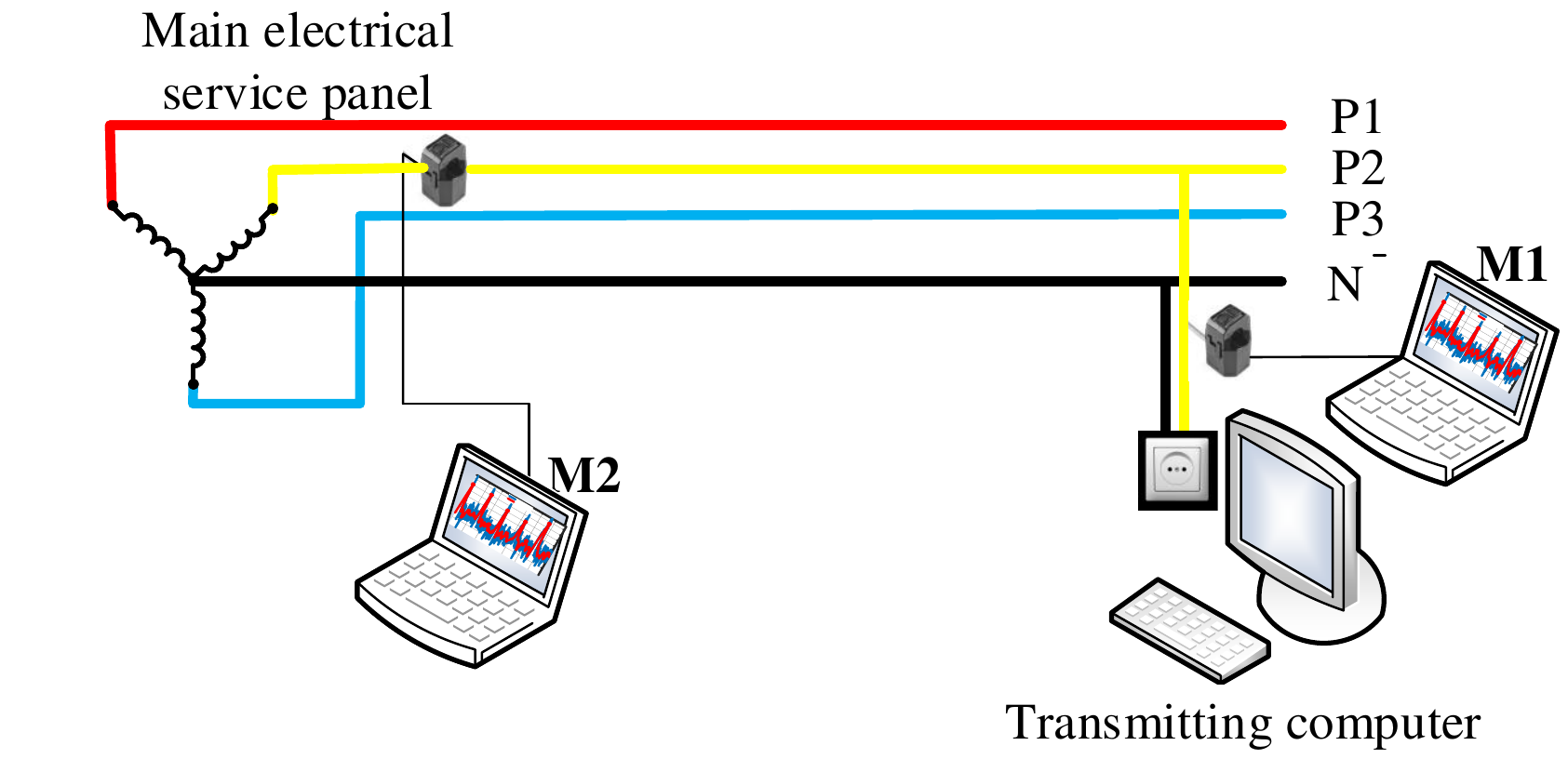}
	\caption{The measurement setup}
	\label{fig:measure_setup}
\end{figure}

In this current sensor the output is sent through a 3.5mm audio cable. We used a laptop computer (Dell Latitude E7450, Windows 10 64-bit) as a measurement system, by connecting the output cable to the audio input jack. The sampling rate was set to 48kHz with a 16-bit resolution. For the signal processing and data analysis we used the MathWorks MATLAB. For real-time demodulation we used a custom FSK demodulator we developed in Visual C++. As can be seen in Fig. \ref{fig:measure_setup} we installed two measurement systems. System M1 was connected to the power line feeding the transmitting computer and used for the evaluation of the line level power-hammering. System M2 was connected to the power line in the main electric panel and used for the evaluation of the phase level power-hammering.

\begin{table*}[]
	\centering
	\caption{The computers used for the transmissions}
	\renewcommand{\arraystretch}{1.2}	
	\label{tab:TableListComp}
	\begin{tabularx}{\textwidth}{|X|X|l|l|l|l|}
		\hline
		\# & Name                            & Model                                                          & Motherboard                                                              & CPU                                                                                                          & PSU                               \\ \hline
		1  & \textbf{PC}                     & \begin{tabular}[c]{@{}l@{}}Silverstone \\ Desktop\end{tabular} & Gigabyte H87M-D3H                                                        & \begin{tabular}[c]{@{}l@{}}Intel Core i7-4770 CPU@ 3.4GHz\\ 4 cores (8 threads)\end{tabular}                 & FSP300-50HMN 300W                     \\ \hline
		2  & \textbf{Server}                 & IBM System x3500 M4                                            & Intel C602J                                                              & \begin{tabular}[c]{@{}l@{}}Intel Xeon CPUE5-2620\\ 12 cores (24 threads)\end{tabular}                        & DPS-750AB-1 A 750Wx2                    \\ \hline
		3  & \textbf{Low power device (IoT)} & \begin{tabular}[c]{@{}l@{}}Raspberry\\ Pi 3\end{tabular}       & \begin{tabular}[c]{@{}l@{}}Raspberry Pi 3\\ \\ Model B V1.2\end{tabular} & \begin{tabular}[c]{@{}l@{}}Quad Core Broadcom\\ BCM2837 64-bit \\ ARMv8,\\ processor Cortex A53\end{tabular} & Stontronics DSA-13PFC-05, 5V 2.5A \\ \hline
	\end{tabularx}
\end{table*}

\subsubsection{Transmitters}
We measured the transmissions from three types of computers: a desktop PC, a server, and a low power device representing an IoT. With the low power device (Raspberry Pi 3), we checked the feasibility of transmitting from IoT devices that only have low power consumption. The list of the computers with their specifications is provided in Table  \ref{tab:TableListComp}. The PC was running a Linux Ubuntu OS
version 16.04.4 LTS \mbox{64-bit}. The server was running a Linux Ubuntu OS version 16.04.1 LTS 64-bit (kernel version 4.4.0). The Raspberry Pi was running the Raspbian Strech OS (kernel version 4.9). The transmitter presented in Section \ref{sec:communication} was compiled with GCC and executed on each of the three computers.

\subsection{Line Level Attack}
In this sub-section we evaluate the line-level attack. We examine the whole spectrum and strength of the generated signal. We also measured the bit error rate and perform further analysis.

\subsubsection{Spectral View}
\begin{figure}
	\centering
	\includegraphics[width=\linewidth]{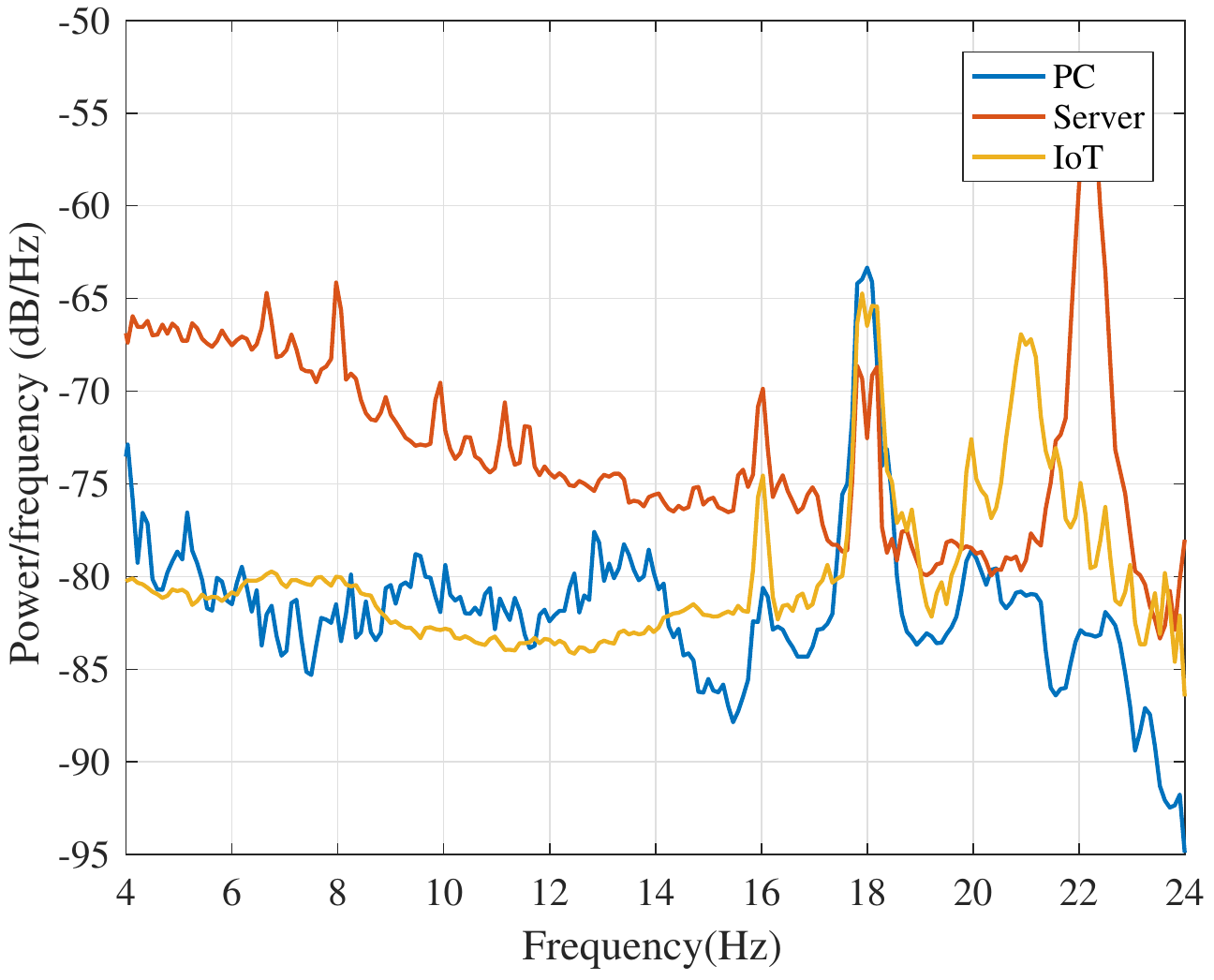}
	\caption{The PSD of the background noise for the PC, server, and IoT device (line level)}
	\label{fig:noise1}
\end{figure}

Fig. \ref{fig:noise1} shows the power spectral density (PSD) of the background noise in a band of 0-24kHz, as measured in the line level. The measurements were taken with the three computers connected to the power line during their normal (non-transmitting) operation. The PSD presents the background noise for each of the three computers. It's noticeable that the frequencies below 5kHz are significantly more noisy than the higher frequencies (5kHz-24kHz). Our experiments show that the transmissions in the 0-5kHz band yield high bit error rates (greater than 25\%) due to the large amount of noise and the low SNR levels. Thus, we consider only the 5kHz-24kHz band for the transmissions in this covert channel. 

\subsubsection{Signal Strength}
\begin{figure}[h!]
	\centering
	\includegraphics[width=\linewidth]{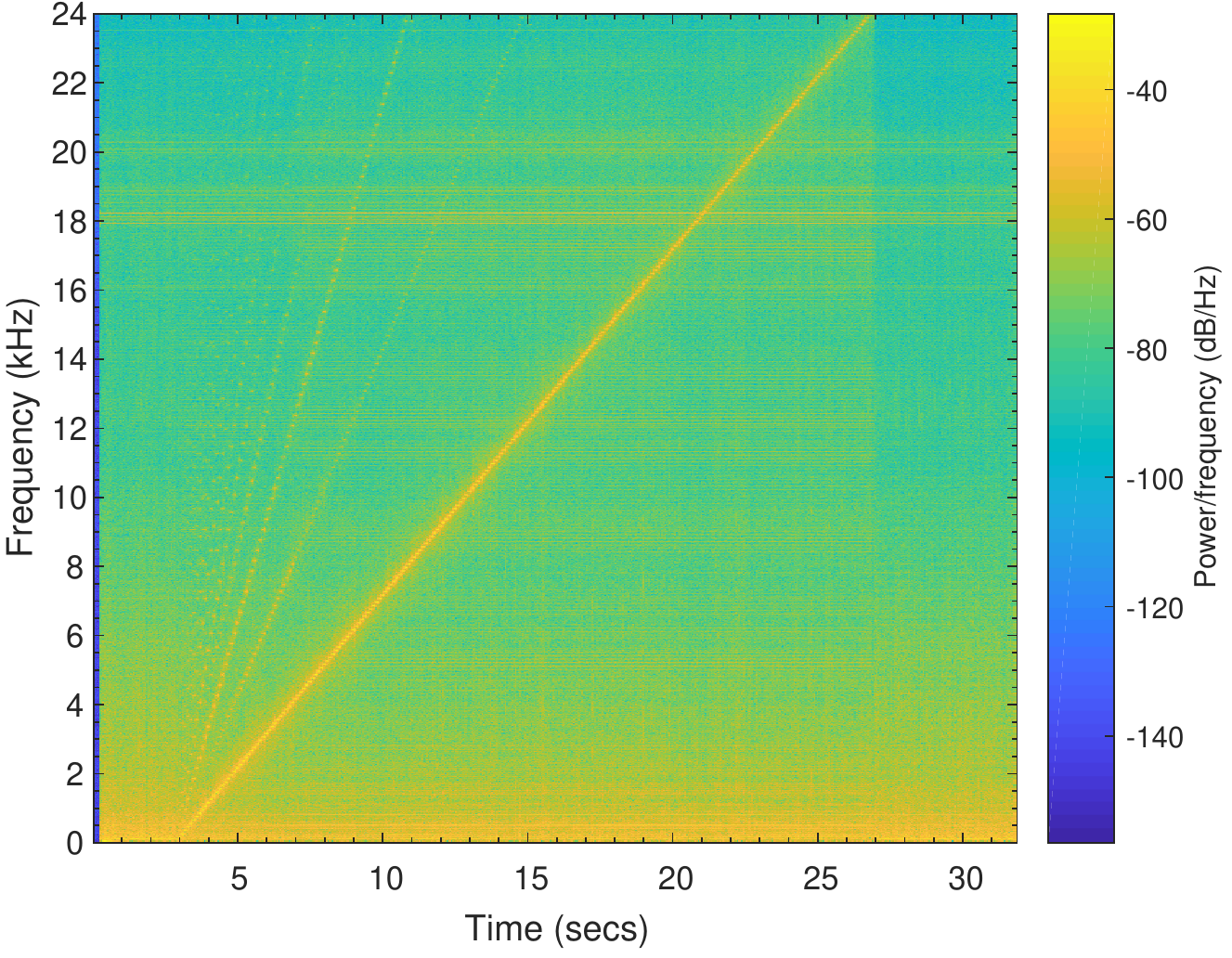}
	\caption{A spectrogram of a sweep signal (line level)}
	\label{fig:pcspectrogramsweep}
\end{figure}

\begin{figure}
	\centering
	\includegraphics[width=\linewidth]{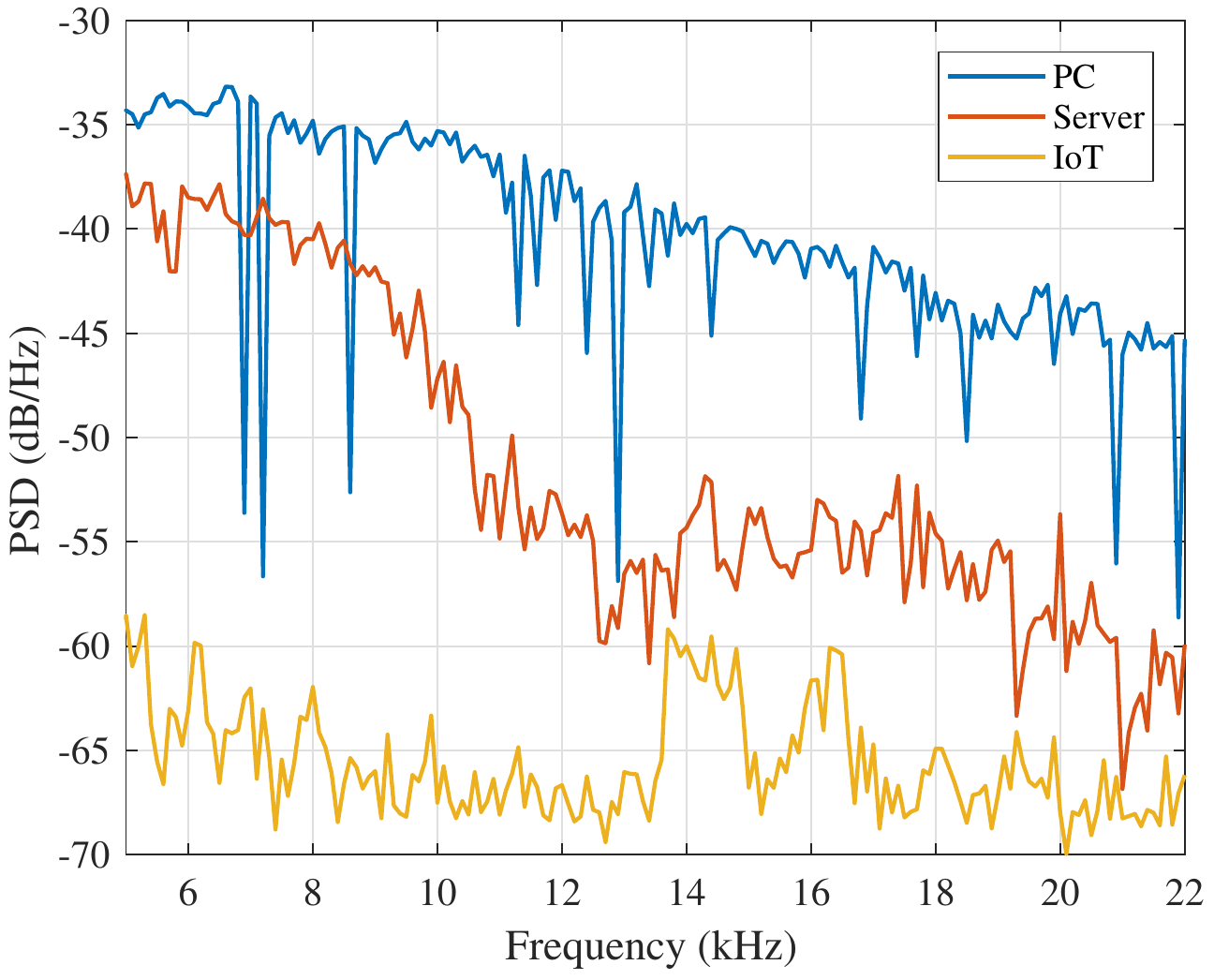}
	\caption{The PSD of a sweep signal transmitted from the PC, server and IoT device}
	\label{fig:LLPSDSweep}
\end{figure}

Figure \ref{fig:pcspectrogramsweep} shows the spectrogram of a sweep signal (from 0 to 24kHz) transmitted from the three computers. Fig. \ref {fig:LLPSDSweep} shows the PSD of the signal. As can be seen, the strongest signal is generated by the PC and the weakest by the IoT. At frequencies above 12kHz, the signal generated by a PC is stronger by 15dB than the server signal and by 25dB than the IoT device signal. The signal power in all three computers decreases with higher frequencies, due to the lower amount of energy per bit.

\subsubsection{Number of Cores}

\begin{figure}[h!]
	\centering
	\includegraphics[width=\linewidth]{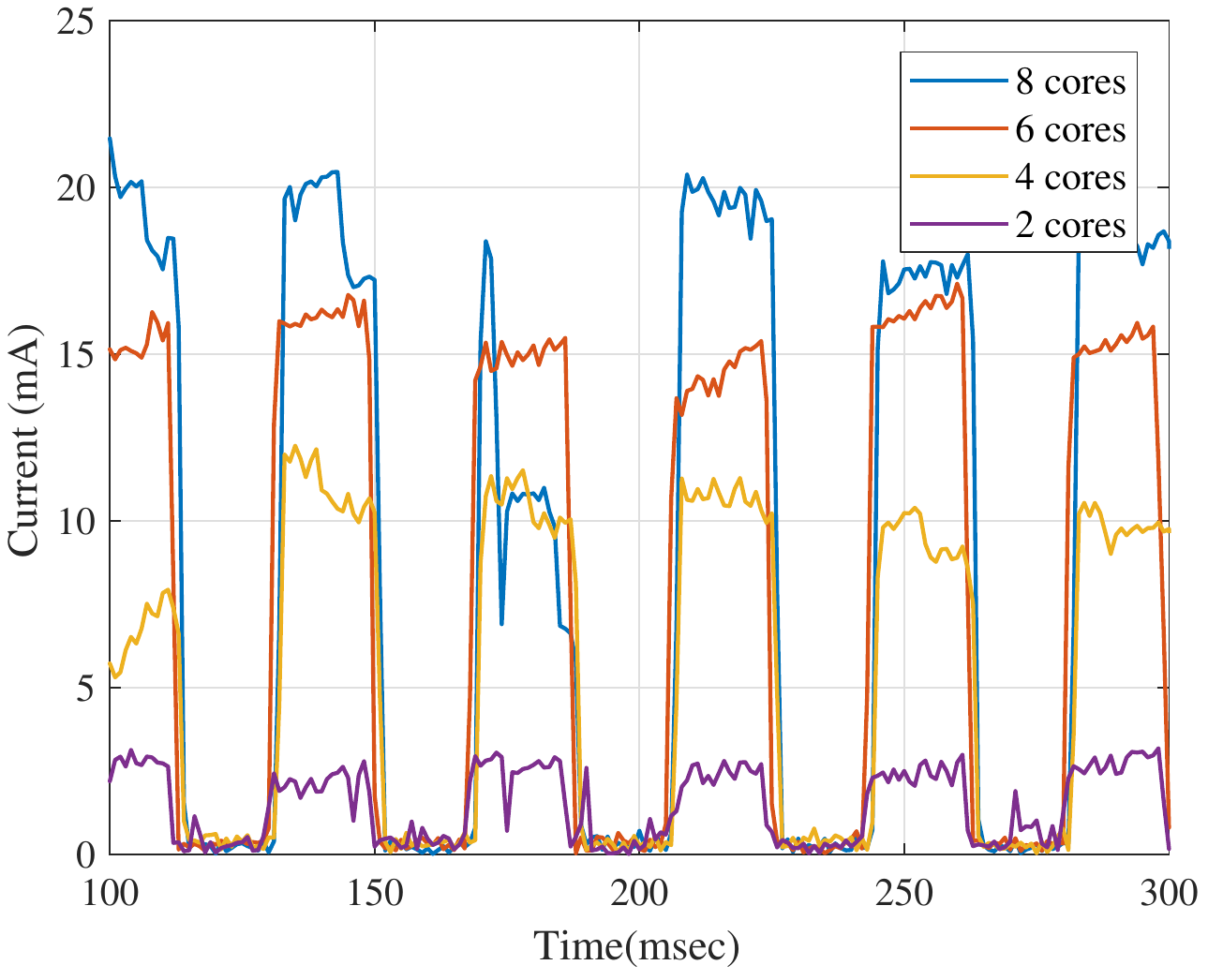}
	\caption{The waveform of a transmission with various number of cores}
	\label{fig:fig_cores}
\end{figure}

The number of cores used in the transmission directly influences the power consumption of the computer, (i.e., more transmitting threads result in greater power consumption), and hence, the amount of conductive emission. The measurements in Figure \ref{fig:fig_cores} show the waveform of an alternating signal ('101010101010') transmitted from the PC at 10kHz. The transmissions in two, four, six, and eight cores yield signal levels of ~2.5mA, ~12mA, ~15mA, and 19mA, respectively. As can be seen, the number of cores used for the transmissions is correlated with the conducted emission measured. However, it is important to note that although using more cores yields a stronger signal, the attacker may use only part of available cores for the transmissions. Using all of the cores will significantly affect the workload on the system and might reveal the malicious activity. 

\subsubsection{Bit Error Rate (BER)}
We measured the bit error rate (BER) for the PC, the server, and the IoT device at various transmission rates. In these experiments we repeatedly transmitted sequences of random bits, decoded them, and compared the results with the original data. The results are summarized in Table \ref{tableber1}.  

\begin{table}[]
	\centering
	\caption{Bit Error Rates (BER)}
	\label{tableber1}
	\renewcommand{\arraystretch}{1.25}	
	\begin{tabular}{|l|l|}
		\hline
		\#     & bit rate (bit error rate)                                                                                                                \\ \hline
		PC     & \textbf{\begin{tabular}[c]{@{}l@{}}333   bit/sec (0\%)\\ 500   bit/sec (0\%)\\ 1000 bit/sec (0\%)\end{tabular}}                      \\ \hline
		Server & \textbf{\begin{tabular}[c]{@{}l@{}}100 bit/sec (0\%)\\ 200 bit/sec (0.96\%)\\ 333 bit/sec (4.8\%)\\ 500 bit/sec (26\%)\end{tabular}} \\ \hline
		IoT    & \textbf{\begin{tabular}[c]{@{}l@{}}5 bit/sec (1.9\%)\\ 10 bit/sec (4.8\%)\\ 20 bit/sec (18.2\%)\end{tabular}}                          \\ \hline
	\end{tabular}
\end{table}

As can be seen, with the PC we maintained bit rates of 333 bit/sec, 500 bit/sec and 1000 bit/sec without errors. With the server we achieved bit rates of 100 bit/sec without errors, 200 bit/sec with less than 1\% errors and 300 bit/sec with 4.8\% errors. The higher bit rate of 500 bit/sec yields an intolerable error rate of 26\%. With the low power Raspberry Pi we could only maintain low bit rates of 1 bit/sec and 10 bit/sec with 1.9\% errors and 4.8\% errors, respectively. For 50 bit/sec we maintained an unacceptable error rate of 18.27\%.

The results indicate that desktop computers and servers could be used to exfiltrate a considerable amount of information such as images, documents and keylogging data. Low power devices are relevant for the exfiltration of small amounts of data such as passwords, credential tokens, encryption keys and so on.     

\subsubsection{Bit Error Rate Analysis}
\begin{figure}
	\centering
	\includegraphics[width=\linewidth]{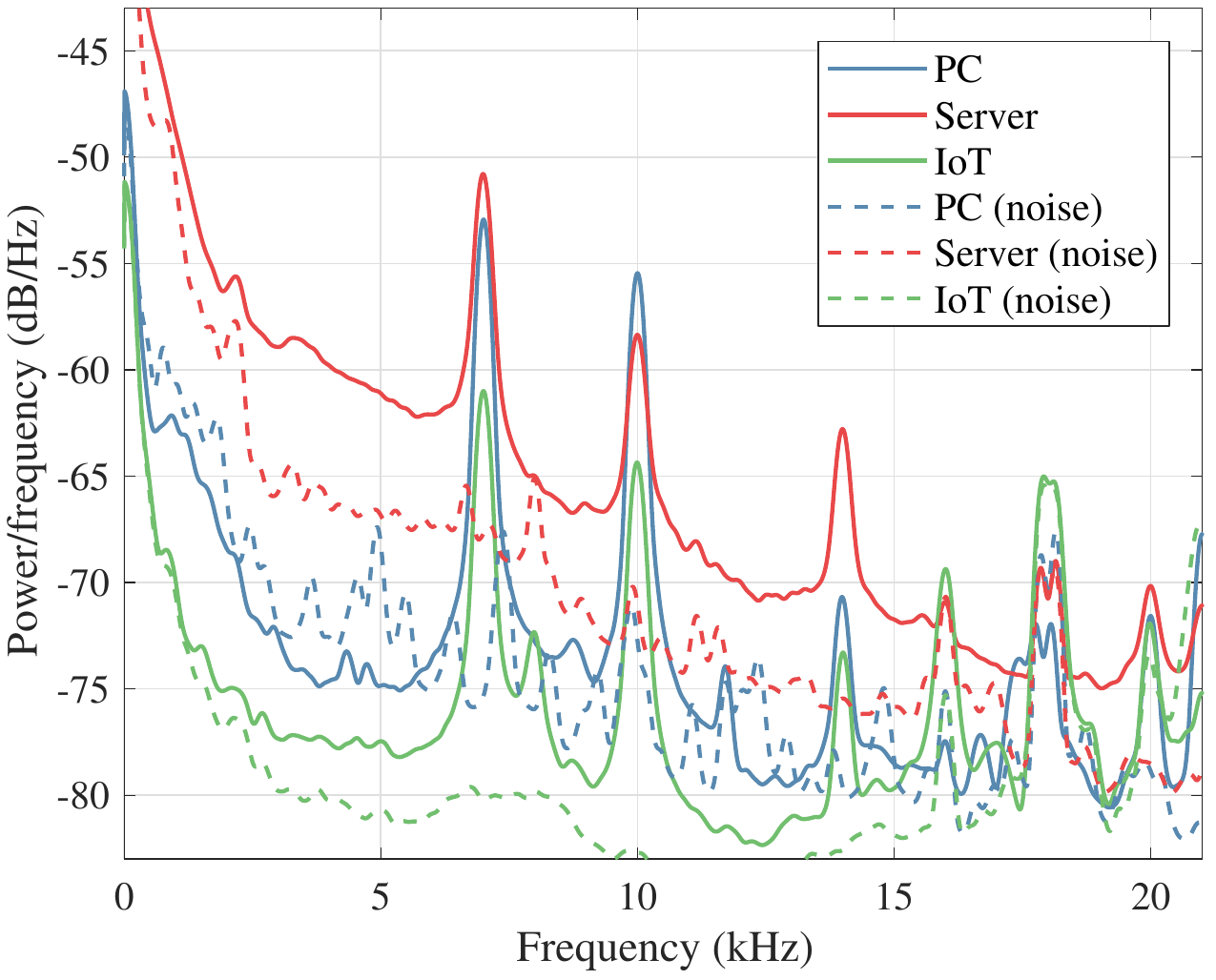}
	\caption{The PSD for PC, server and IoT device (B-FSK with 7Khz and 10kHz carriers)}
	\label{fig:plot_psd_compare}
\end{figure}

In order to maintain a reliable channel with low BER levels, the less noisy frequencies must be chosen. We analyze the theoretical BER limit where data was transmitted in a quiet frequency region. For each of the three computers (PC, server, and IoT device) we calculated the PSD with and without the presence of the signal. In this case, the data was transmitted in B-FSK at two quiet frequencies: 7kHz and 10kHz. The PSD in Fig. \ref{fig:plot_psd_compare} shows how the power of a signal is distributed in a range of 0-24kHz. With regard to our communication scheme (a non-coherent B-FSK), the theory requires a noise margin of 11dB to maintain a BER of $10^{-3}$ and a noise margin of 8dB to maintain a BER of $2\times 10^{-2}$  \cite{Proakis2007}.
% \cite[Sec. 5-4-4]{Proakis2007_4th}.\, pp. 394
%\cite[Eq. (5-4-53)]{Proakis2007_4th}
%
The PSD in Fig. \ref{fig:plot_psd_compare} shows that a noise margin of approximately 13dB is measured for the transmissions in the quiet frequencies (7kHz and 10kHz). This indicates that in this communication scheme it is possible to establish a reliable channel while maintaining relatively low levels of BER. Note that in a real world attack, it's not always possible to select the most quiet frequencies, since the spectral view in the target is unknown to the attacker in advance. 

\subsection{Phase Level Attack}
In this sub-section we evaluate the phase level attack. We examine the whole spectrum and strength of the generated signal. We also measured the bit error rate. 
Note that background noise at the phase level is significantly higher than the noise in the line feeding the computer. This is due to the fact that all of the consumers share the power line; computers, TVs, lights, electronic devices, and so on.

\subsubsection{Spectral View}

\begin{figure}[h!]
\centering
\includegraphics[width=\linewidth]{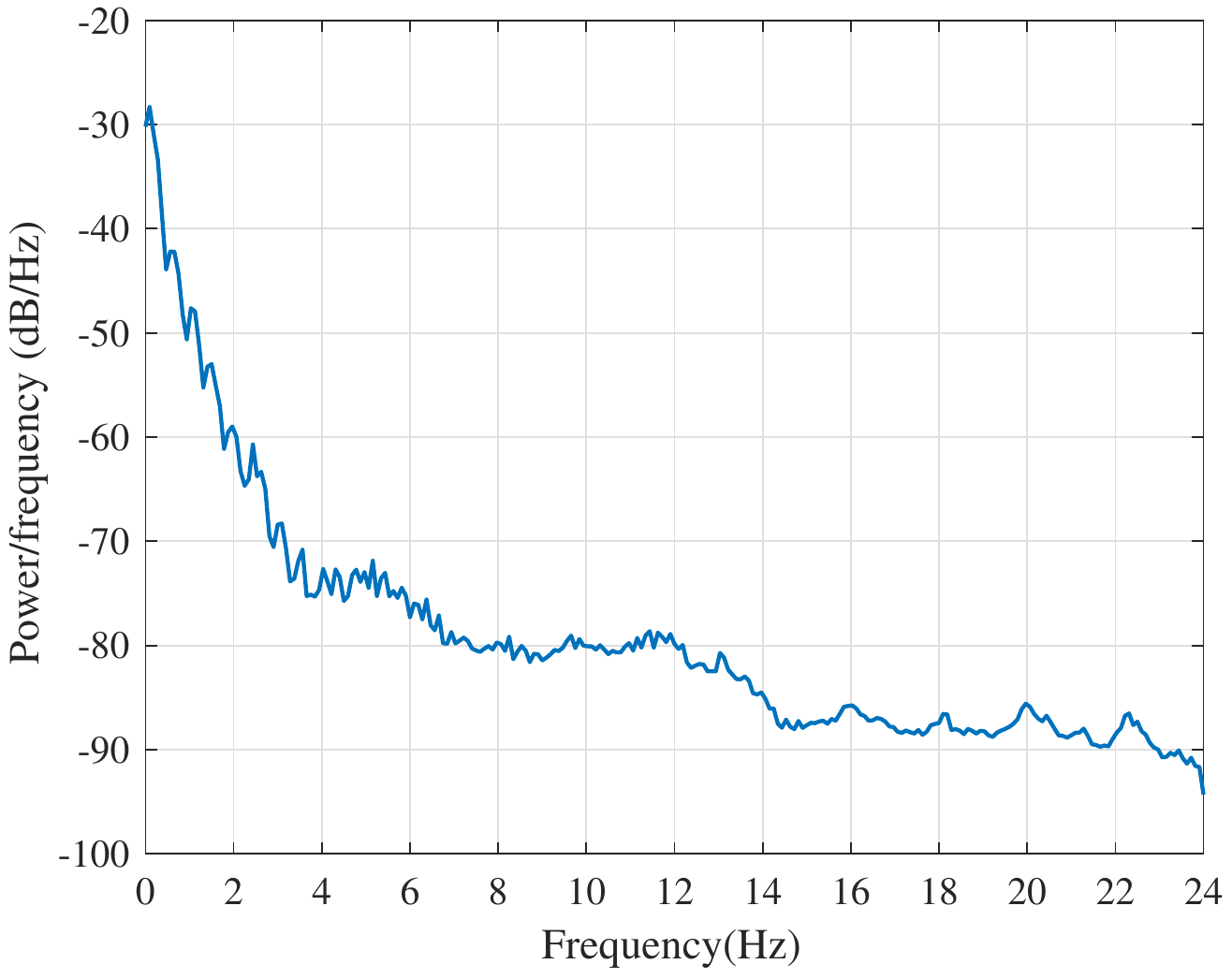}
\caption{The PSD of the background noise (phase level)}
\label{fig:gridnoisePSD}
\end{figure}

Fig. \ref{fig:gridnoisePSD} shows the PSD of the background noise in a band of 0-24kHz, as measured in the phase level. The PSD represents the typical background noise in this frequency band. As can be seen, the 15kHz-24kHz band is less noisy than the 0-15kHz band.  
It is important to note that a spectral view in the phase level is highly dependents on the specific environment, and particularity on the amount (and types) of the electric equipment connected to the power lines. We considered our lab environment as representative for IT environments in this regard. The lab consists of various electrical consumers: several desktop PCs, networking equipment (network switches and routers, Wi-Fi), office lighting, a TV, and various types of electronic equipment.

\subsubsection{Signal Strength}
\begin{figure}[h!]
	\centering
	\includegraphics[width=1\linewidth]{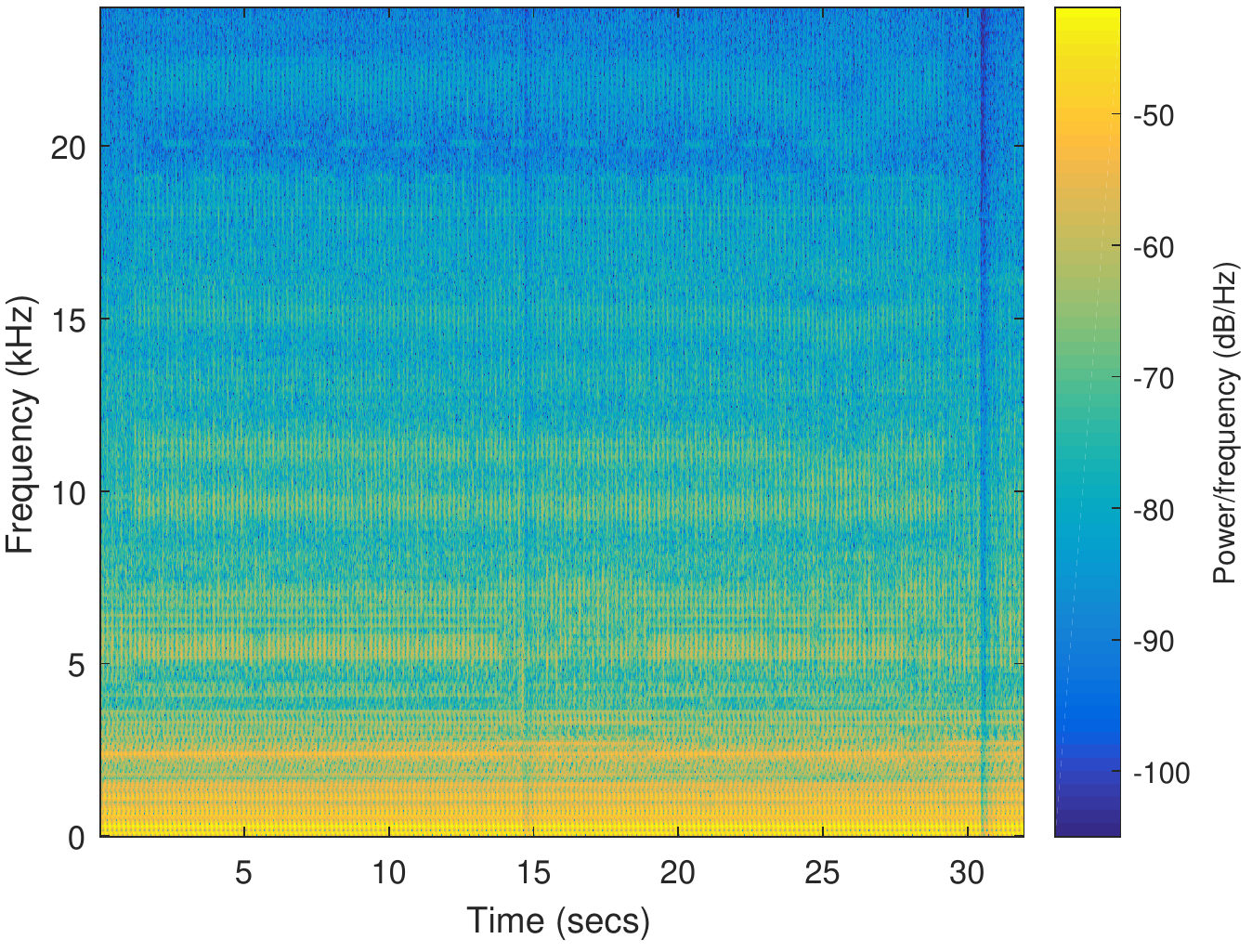}
	\caption{B-FSK transmission in 19kHz and 20kHz}
	\label{fig:griddata}
\end{figure}

The spectrogram in Fig. \ref{fig:griddata} presents a binary sequence ('1010101010') transmitted from the PC as measured in the phase level. The signal amplitudes at a high value (logical '1') is 0.3mA and at a low value (logical '0') is 0.15mA (a difference of 6dB power in terms of SNR). Notably, the measured signals are significantly weaker than the signals measured at the line level, because of the greater amount of noise in the phase level.

\subsubsection{Bit Error Rate (BER)}
We measured the bit error rate for  four bit rates, with the transmitting PC. As in the line level BER calculations, we transmitted sequences of 200 random bits, decoded them, and compared the results with the original data. The tests were repeated three times. The results are summarized in Table \ref{tableber2}.  

\begin{table}[]
	\centering
	\caption{Bit-Error-Rates (BER)}
	\label{tableber2}
	\renewcommand{\arraystretch}{1.25}	
	\begin{tabular}{|l|l|}
		\hline
		\# & bit rate (bit error rate)                                                                                 \\ \hline
		PC & \textbf{\begin{tabular}[c]{@{}l@{}}1 bit/sec (0\%)\\ 3 bit/sec (0\%)\\ 5 bit/sec (1\%) \\ 10 bit/sec (4.2\%) \end{tabular}} \\ \hline
	\end{tabular}
\end{table}

Due to the noise in the phase level, we could only achieve a sustainable BER with low bit rates. We maintained bit rates of 1 bit/sec, 3 bit/sec and 5 bit/sec with 0\%, 0\%, and 1\% errors, respectively. With 10 bit/sec we measured 4.2\% errors. The results indicate that in the phase level power-hammering attack, desktop computers could only be used to exfiltrate small amount of data such as passwords, credential tokens, encryption keys, and so on.

\subsection{Virtual Machines (VMs)}
Virtualization technologies are widely used in modern IT environments. In the context of security, one of the advantages of virtualization is the resource isolation it provides. Virtual machine monitors (VMM) provide a layer of abstraction between the virtual machine and the physical hardware (CPU and peripherals). Since the signal generated is tightly related to the CPU timing and performance, we examined whether the virtualization layer caused interruptions or delays which may affect the signal quality.

\begin{figure}
	\centering
	\includegraphics[width=\linewidth]{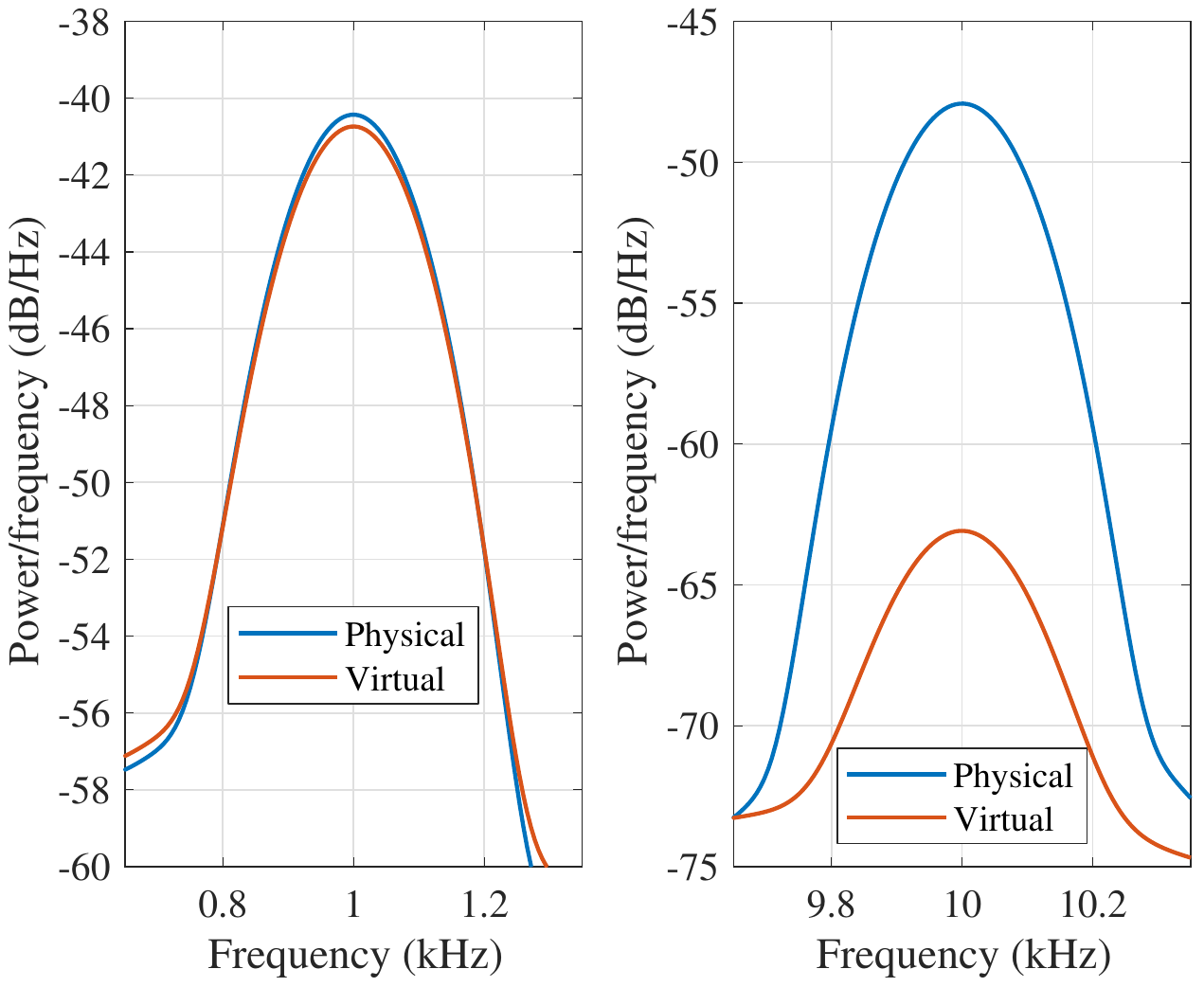}
	\caption{The signals generated from physical and virtual machines at 1kHz and 10kHz}
	\label{fig:plot_virtualization_diff}
\end{figure}

In order to evaluate the effect of the VM, we compared the signals generated from a physical machine (host) and the signals generated from a VM (guest). In these measurements, the host and the guest were running Linux Ubuntu 16.04 64-bit. We used VMWare Workstation Player 14.0 and configured the guest to support four processors. For the comparisons, an identical sweep signal in the range of 0-24kHz was transmitted from both the host and guest. The analysis of the signals received showed that while the host was able to generate signals in the whole frequency range of 0-24kHz, the guest could only generate signals up to 7kHz. Fig. \ref{fig:plot_virtualization_diff} shows the PSD of transmissions made by host and guest machines with two different frequencies (1kHz and 10kHz). As can be seen,  
the 1kHz signal is similar in the host and guest (with levels of -41dB). The 10kHz guest signal (-63dB) is significantly weaker than the host signal (-47dB).

The 7kHz limitation shown above is a result of the interruptions occurring in the transmitting threads when executed in a VM. Technically, the VMM initiates a periodical context switch which suspends the transmitting process (and its threads), in order to transfer the control to the host machine. These interruptions effectively limit the operational frequency of the transmitting threads to 7kHz.    

\subsection{Interference}

\begin{figure}[b]
	\centering
	\includegraphics[width=\linewidth]{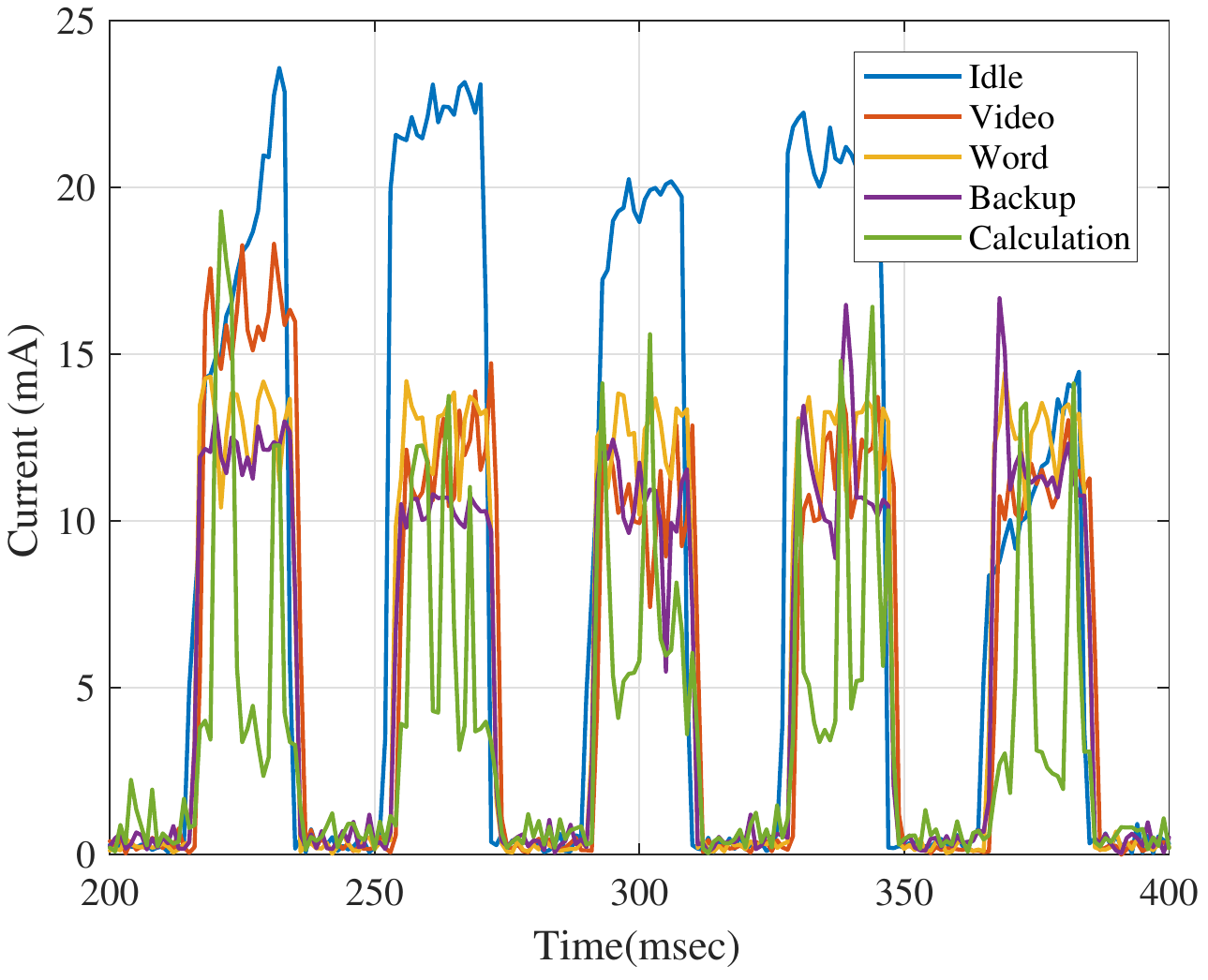}
	\caption{Transmissions with interfering processes}
	\label{fig:interfer}
\end{figure}
Recall that we generate the signals by regulating the workloads on the CPU. Since the transmitting process (and its theads) shares the CPU with other processes in the operating system, we examined whether the activity of other processes interfere with the signal generation. For the evaluation, we tested the following workloads on the system.

\begin{enumerate}
	\item 	\textbf{Idle}. The system is idle and only the default processes are running in the background.
	\item 	\textbf{Word processing}.  The LibreOffice Writer \cite{HomeLibr6:online} is open, and the user is typing a document.
	\item 	\textbf{Video playing}. The VLC media player \cite{Official45:online} is playing an HD video clip.
	\item \textbf{Backup}. The Linux \texttt{rsync} \cite{rsync1Li64:online} command is performing a backup to the HDD.
	\item \textbf{CPU intensive calculations}. The Linux matho-primes \cite{UbuntuMa29:online} is performing  calculations of big prime numbers.
\end{enumerate}

In these tests, we transmitted an  alternating binary sequence ('1010101') in the B-FSK modulation, in the frequency band of 16kHz-19kHz. The signals were transmitted from the PC and measured at the line level. The measurements in Fig. \ref{fig:plot_virtualization_diff} show that the covert channel is usable even when other active processes are running in the system.  The idle state where no other processes interfered with the transmitting process yielded the strongest signal of more than 20mA. Operations like video playing, word processing and backup, decrease the signal strength by 25\% to levels of 10-15mA.
The CPU intensive operations (like calculations) add a significant amount of noise to the generated signal, dramatically increasing the bit error rate. 

\subsection{Multiple Frequency-Shift Keying (M-FSK)}
\begin{figure}[h]
	\centering
	\includegraphics[width=0.8\linewidth,page=1]{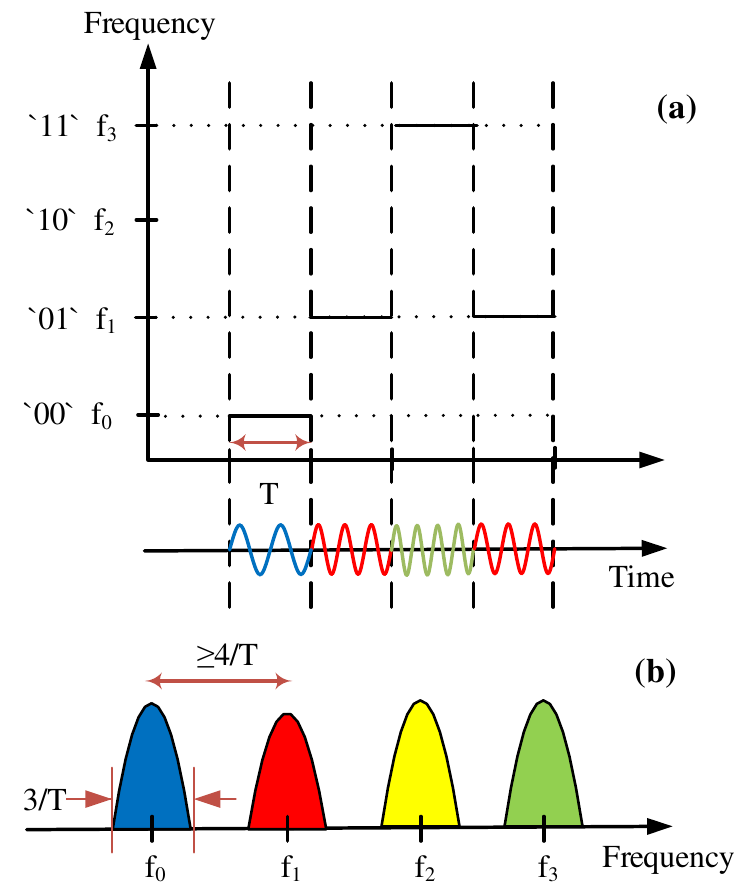}
	\caption{Illustration of the M-FSK modulation}
	\label{fig:fsk_illustration}
\end{figure}

We improved the communication performance by applying M-FSK modulation instead of B-FSK. In this modulation, $M$ different carrier frequencies are used instead of two carriers in a case of B-FSK. Each carrier frequency is encoding a different symbol in the alphabet, as illustrated in Fig. \ref{fig:fsk_illustration}a. Typically, each symbol represents $\log_2(M)$ bits.

The resulting bandwidth, $B$, is illustrated in Fig. \ref{fig:fsk_illustration}b and is given by 
\begin{equation}
B = M\Delta f + B_c\cong\frac{4M+3}{T},
\end{equation}
where
$B_c = 3/T$ is the bandwidth of each modulated carrier and $\Delta f=4/T$ is the minimum spacing between modulation frequencies.
The resulting bit rate is given by
\begin{equation}
R = \frac{\log_2(M)}{T}.
\end{equation}

\begin{figure}[h!]
	\centering
	\includegraphics[width=\linewidth]{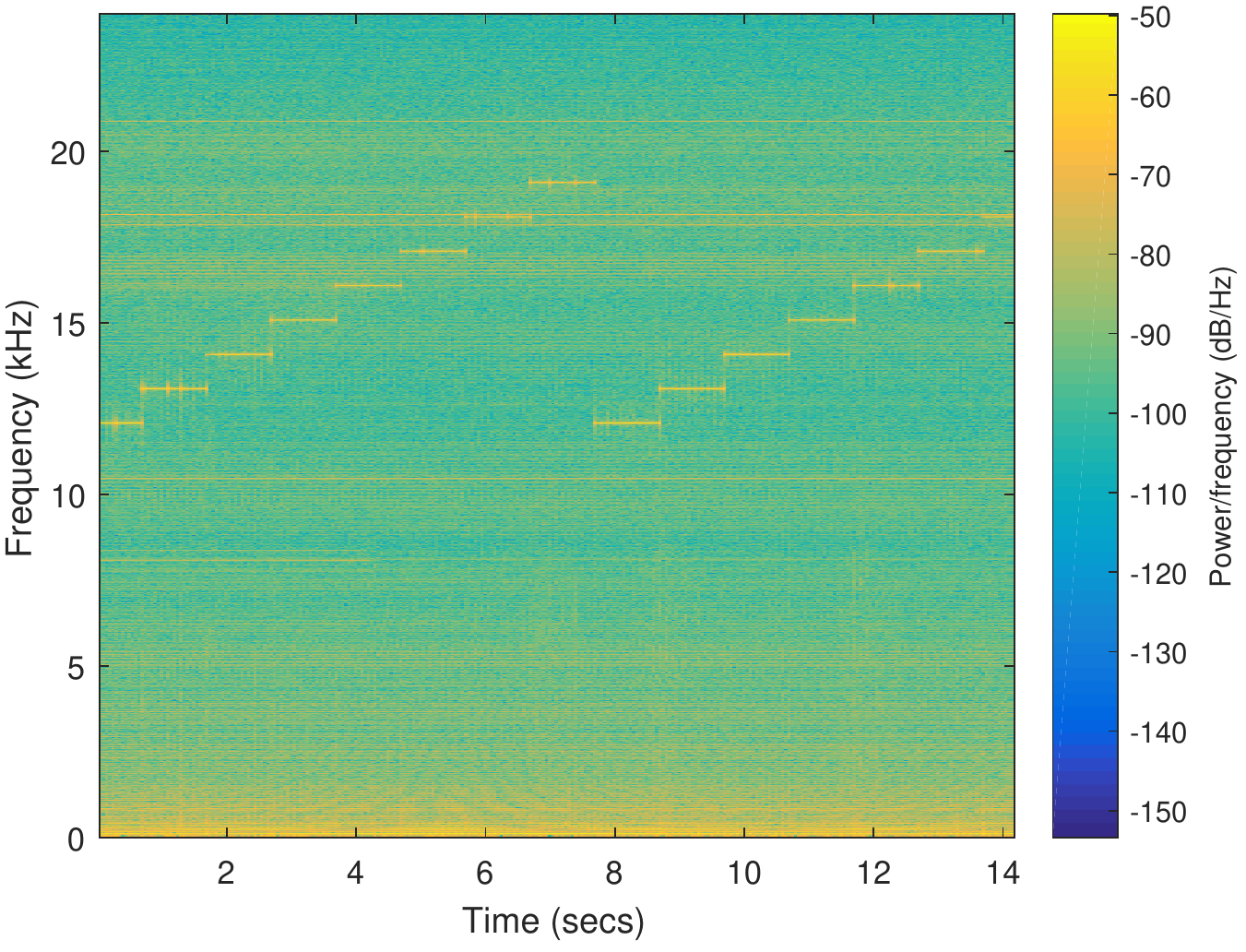}
	\caption{Spectrogram of 8-FSK transmission}
	\label{fig:MFSK}
\end{figure}

Fig. \ref{fig:MFSK} shows the spectrogram of a transmission in 8-FSK modulation in the 12kHz-18kHz band, measured at the line level. Recall that we successfully tested transmission at 1000 bit/sec in B-FSK ($T=0.001$). In the case of 8-FSK modulation ($M=8$), the bandwidth is increased to $R=4000$ bit/sec.   

\subsection{Attenuation}
In power line communication (PLC), the quality of a signal attenuates  with the distance it travels on the wires. The signal attenuation reported for low power (in door) electrical networks and the frequencies of interest in our covert channel is approximately 10dB/km \cite{Ferreira2010}. Therefore, in the line level and phase level attacks, the signal quality is affected mainly by the noise in the electrical network and less by attenuation.

\subsection{Stealth and evasion}
The transmitting program only leaves a small footprint in the memory, making its presence easier to hide from intrusion detection systems. At the OS level, the transmitting program requires no special or elevated privileges (e.g., root or admin), and hence can be initiated from an ordinary user space process. The transmitting code consists mainly of basic CPU operations such as busy loops, which do not expose malicious behaviors, making it highly evasive from automated analysis tools.

\section{Countermeasures}
\label{sec:counter}

\subsubsection{Power line monitoring}
A primary defensive countermeasure in the field of covert channels is monitoring the channel in order to detect the presence of covert communication \cite{zander2007survey}. In our case, it is possible to detect the covert transmissions by monitoring the currency flow on the power lines. The measurements are continuously analyzed to find hidden transmission patterns or deviations from the standard behavior. The monitoring can be done in a non-invasive way using the split core current transformer used for the attack. Note that detecting covert transmissions in the noisy spectrum is known to be a challenging task and may not bring reliable results \cite{carrara2016air,goher2012covert}.

\subsubsection{Signal filtering} 
Power line filters are aimed at limiting the leakage of conduction and radiation noise to the power lines \cite{counterliu2002high,countershih1996procedure,counterlin1994reduction,counterye2004novel}. Such filters, also known as EMI filters, are primarily designed for safety purpose, since noise generated by a device in the power network can affect other devices, causing them to malfunction. As a preventive countermeasure, it is possible to limit the signal generated by our covert channel by attaching
filters to the power lines in the main electrical service panel. Note that in order to prevent the line level power-hammering attack, such filters must be installed at every power outlet. This is possible by placing filters between the power supply of the computers and the power sockets or by integrating them into the power-supply itself \cite{wolmarans2002technology}.  
It is important to note, however, that most of the filters designed for limiting the conducted emission are for higher frequencies \cite{AN2162Si53:online}. As shown in the evaluation, our covert channel  enables transmissions at frequencies lower than 24kHz, hence bypassing most of the filters.
Note that there are various regulations about the levels of radiated and conducted emission leaked from power supplies. For example, the FCC Part 15 certification requires that the conducted emission be controlled in the 450kHz-30MHz frequency band. \cite{Microsof90:online}. However, our covert channel utilizes a much lower frequency band (0-24kHz), and hence works even with equipment that is fully complaint with the regulations.

\subsubsection{Signal jamming}
A  software level jamming solution involves the execution of background processes that initiate random workloads on the system. The random signals interfere with the transmissions of the malicious process. The main limit of this approach is that the random workloads weaken system performance and may be infeasible in some environments (e.g., real-time systems). Jammers can also be implemented in a form of hardware component. In this approach, a dedicated electronic component employ random loads on the power lines, masking the signals generated by other devices. Note that the hardware jammer solution is only relevant for the phase level attack (since the jammer device is installed on a phase level), and it is not effective against the line-level power-hammering. 

\subsubsection{Host-Based Detection} 
In this approach, host based intrusion detection systems (HIDSs) and host based intrusion prevention systems (HIPSs) continuously trace the activities of running processes in order to detect suspicious behavior; in our case, a group of threads that abnormally regulates the CPU workloads would be reported and inspected. Such a detection approach would likely suffer from a high rate of false alarms, since many legitimate processes use CPU intensive calculations that affect the processor’s workload. Another problem in the runtime detection approach is that the signal generation algorithm (presented in Section \ref{sec:communication}) involves only non-privileged CPU instructions (e.g., busy loops). Monitoring non-privileged CPU instructions at runtime necessitates that entering the monitored processes enter a so-called step-by-step mode, which severely degrades performance \cite{guri2015gsmem}. Finally, software based detection also suffers from an inherent weakness in that it can easily be bypassed by malware using evasion techniques. In our case, the malware may inject the transmitting threads into a legitimate, trusted (signed) process to bypass the security mechanisms.

\section{Conclusion}
\label{sec:conclusion}
In this paper we present \textit{PowerHammer}, a new type of attack that uses the power lines to exfiltrate data from air-gapped computers. We implemented a malicious code that controls the power consumption of a computer by regulating the momentary utilization of the CPU cores. Data is modulated in the fluctuations of the power consumption, and then conducted from the power supply to the power lines. In \textit{line level power-hammering}, the attacker 
places a probe on the the power cables feeding the computer. In \textit{phase level power-hammering} the probe is placed in the main electrical service panel of the whole floor. The attacker measures the conducted emission on the power lines, processes the signal and decodes it back to binary information. We evaluated the covert channel in different scenarios with three types of computers: a desktop PC, a server and a low power IoT device. We checked the effect of various forms of interference and the use of virtual machines on the transmissions. We also examined detection and prevention countermeasures. The results show that data can be exfiltrated from air-gapped computers through the power lines at bit rates of 1000 bit/sec for line level power-hammering, and 10 bit/sec for phase level power-hammering.

\bibliographystyle{ieeetran}
\bibliography{PowerHammer,cyber_dima}

\end{document}